\RequirePackage{silence}
\documentclass[10pt,aps,pra,twocolumn,showpacs,superscriptaddress,nofootinbib]{revtex4-2}
\usepackage[english]{babel} 
\usepackage[usenames, dvipsnames]{color}
\usepackage{graphicx}
\usepackage{bm}
\usepackage{mathtools}
\usepackage{amsthm} 
\usepackage{amssymb} 
\usepackage{times} 
\usepackage[pdftex,colorlinks=true,urlcolor=Blue,linkcolor=Blue,citecolor=RedViolet]{hyperref}
\usepackage{newpxtext} % fonte texto

\newcommand{\hy}[1]{\hyperlink{#1}{\color{Gray} #1}}
\newcommand{\ii}{\mathsf{i}}
\newcommand{\tr}{{\rm Tr}}
\newcommand{\blue}[1]{{#1}}
\begin{document}
\title{%Parameter Tuning for 
Complete Positivity and Thermal Relaxation in Quadratic Quantum Master Equations}
\author{F. Nicacio}
\email{nicacio@if.ufrj.br} 
\affiliation{Instituto de Física, 
             Universidade Federal do Rio de Janeiro, 
             Rio de Janeiro, RJ 21941-972, Brazil}
\author{T. Koide}
\email{koide@if.ufrj.br}
\affiliation{Instituto de Física, 
             Universidade Federal do Rio de Janeiro, 
             Rio de Janeiro, RJ 21941-972, Brazil}  
\affiliation{Frankfurt Institute for Advanced Studies (FIAS), Frankfurt am Main, Germany}           
\date{\today}
%%%%%%%%%%%%%%%%%%%%%%%%%%%%%%%%%%%%%%%%%%%%%%%%%%%%%%%%%%%%%%%%%%%%%%%%%%%%%%%%%%%%%%%%%
%%%%%%%%%%%%%%%%%%%%%%%%%%%%%%%%%%%%%%%%%%%%%%%%%%%%%%%%%%%%%%%%%%%%%%%%%%%%%%%%%%%%%%%%%
\begin{abstract}
The ultimate goal of this project is to develop a systematic method for deriving 
quantum master equations that satisfy the requirements of a 
completely positive and trace-preserving (CPTP) map, further describing thermal 
relaxation processes. 
In this paper, we assume that the quantum master equation is obtained 
through the canonical quantization of the generalized Brownian motion proposed in
[\href{https://doi.org/10.1016/j.physleta.2023.129277}
      {Physics Letters A, {\bf 494}, 129277 (2024)}]. 
At least classically, this dynamics describes the thermal relaxation process 
regardless of the choice of the system Hamiltonian.   
The remaining task is to identify the parameters ensuring that
the quantum master equation meets complete positivity.
We limit our discussion to many-body quadratic Hamiltonians and 
establish a \hy{CPTP} criterion for our quantum master equation.
This criterion is useful for applying our quantum master equation
to models with interaction such as a network model,
which has been used to  investigate how quantum effects modify heat conduction.
\end{abstract}
%%%%%%%%%%%%%%%%%%%%%%%%%%%%%%%%%%%%%%%%%%%%%%%%%%%%%%%%%%%%%%%%%%%%%%%%%%%%%%%%%%%%%%%%%
%%%%%%%%%%%%%%%%%%%%%%%%%%%%%%%%%%%%%%%%%%%%%%%%%%%%%%%%%%%%%%%%%%%%%%%%%%%%%%%%%%%%%%%%%

\maketitle

%%%%%%%%%%%%%%%%%%%%%%%%%%%%%%%%%%%%%%%%%%%%%%%%%%%%%%%%%%%%%%%%%%%%%%%%%%%%%%%%%%%%%%%%%
%%%%%%%%%%%%%%%%%%%%%%%%%%%%%%%%%%%%%%%%%%%%%%%%%%%%%%%%%%%%%%%%%%%%%%%%%%%%%%%%%%%%%%%%%
\section{ Introduction }  %%%%%%%%%%%%%%%%%%%%%%%%%%%%%%%%%%%%%%%%%%%%%%%%%%%%%%%%%%%%%%%
%%%%%%%%%%%%%%%%%%%%%%%%%%%%%%%%%%%%%%%%%%%%%%%%%%%%%%%%%%%%%%%%%%%%%%%%%%%%%%%%%%%%%%%%%
%%%%%%%%%%%%%%%%%%%%%%%%%%%%%%%%%%%%%%%%%%%%%%%%%%%%%%%%%%%%%%%%%%%%%%%%%%%%%%%%%%%%%%%%%
In recent years, there has been considerable discussion about how thermodynamical 
descriptions can be extended to small fluctuating systems.
For this, there are at least two approaches: 
stochastic thermodynamics \cite{sekimoto,seifert,peliti} and 
quantum thermodynamics \cite{qt1,qt2}.
The former considers classical systems and assumes stochastic dynamics 
like Brownian motion and the latter considers 
open quantum systems.

%\red{Compared to classical systems, our understanding of quantum dissipative dynamics 
%is quite limited \cite{dieter}.}
In classical systems, it is possible to introduce a model of Brownian motion 
applicable to any particle Hamiltonian, allowing for the definitions of heat, work, 
and entropy in a manner consistent with thermodynamics \cite{KoiNic2023}.
In open quantum systems, the time evolution is often assumed to be given by
a completely positive and trace-preserving \hypertarget{CPTP}{(CPTP)} map
with the Gorini-Kossakowski-Sudarshan-Lindblad \hypertarget{GKSL}{(GKSL)} equation 
exemplifying such dynamics.
Admittedly heat, work, and entropy can be defined in line with thermodynamics 
through the GKSL equation, at least when the system's Hamiltonian is given 
by a harmonic oscillator, but its applicability
to general Hamiltonians remains unclear.
It is thus still worthwhile to study the systematic derivation of a master equation 
applicable to describe thermal relaxation processes.

Recently, the present authors introduced the generalized model of Brownian motion 
and considered its canonical quantization \cite{KoiNic2023}. 
Compared to the \hy{GKSL} equation, this approach has at least two advantages. 
One is its simplicity: 
our master equation is determined only by the system Hamiltonian and  
a few parameters (coupling constants and temperatures).
The other is that, regardless of the choice of the system Hamiltonian, 
the classical limit of the quantum master equation always describes 
the thermal relaxation process, as we will show.
Noteworthy, besides the asymptotic approaching towards an equilibrium state, 
thermal relaxation here also implies the appropriate inequality 
involving heat and entropy change during the process, 
as dictated by the second law of thermodynamics.
%
% \red{It should be noted that here the term "describes thermal relaxation process" 
%not only refers to the system asymptotically approaching thermal equilibrium or 
%a steady state but also implies that the appropriate inequality between 
%entropy change and heat, as dictated by the second law of thermodynamics, 
%is satisfied during this process.}
%
Nevertheless, the obtained evolution is generally non-\hy{CPTP}.
Interestingly, however, 
for a harmonic oscillator a tuning of
the coupling parameters ensures the conformity with a \hy{CPTP} map \cite{KoiNic2023}.
Therefore, it is important to investigate whether 
our quantum master equation satisfies the \hy{CPTP} requirement
for arbitrary Hamiltonians.

In this paper, we assume that the quantum dynamics is governed
by a master equation obtained through
the canonical quantization of generalized Brownian motion \cite{KoiNic2023}, 
and apply it to a general many-body quadratic Hamiltonian. 
We then establish a \hy{CPTP} map in our quantum master equation
providing a necessary and sufficient criterion,
which represents quantum constraints for physically admissible parameters 
of a genuine quantum dynamics, like an uncertainty principle.
Although the discussion is limited only to quadratic Hamiltonians,
this simplified situation remains of interest for discussing heat conduction, 
where the system Hamiltonian is represented by linearly interacting harmonic oscillators, 
known as the network model.

This paper is organized as follows.
The derivation of the new quantum master equation proposed in Ref.\ \cite{KoiNic2023} 
is briefly summarized in Sec.\ \ref{sec:CQoGBM} 
and its classical limit is derived in Sec.\ \ref{sec:CLotQE} 
\blue{for a general Hamiltonian}. 
This quantum master equation is mapped to a general quadratic form, 
which is called the quadratic time-convolutionless master equation
and then the criterion to obtain a \hy{CPTP}
evolution is discussed in Sec.\ \ref{sec:GKSLME}.
The relationship between our equation and the 
Bloch-Redfield equation is examined in Sec.\ \ref{sec:AtRE}. 
The derived criterion is applied to one degree
of freedom systems in Sec.\ \ref{sec:LKE},
while in Sec.\ \ref{sec:IO}, it is applied to the network model. 
Section \ref{sec:CRaP} is devoted to concluding remarks.

%%%%%%%%%%%%%%%%%%%%%%%%%%%%%%%%%%%%%%%%%%%%%%%%%%%%%%%%%%%%%%%%%%%%%%%%%%%%%%%%%%%%%%%%%
%%%%%%%%%%%%%%%%%%%%%%%%%%%%%%%%%%%%%%%%%%%%%%%%%%%%%%%%%%%%%%%%%%%%%%%%%%%%%%%%%%%%%%%%%
\section{Canonical quantization of generalized Brownian motion} \label{sec:CQoGBM} %%%%%%
%%%%%%%%%%%%%%%%%%%%%%%%%%%%%%%%%%%%%%%%%%%%%%%%%%%%%%%%%%%%%%%%%%%%%%%%%%%%%%%%%%%%%%%%%
%%%%%%%%%%%%%%%%%%%%%%%%%%%%%%%%%%%%%%%%%%%%%%%%%%%%%%%%%%%%%%%%%%%%%%%%%%%%%%%%%%%%%%%%%
In this section, 
we give a brief summary of the derivation of the quantum master equation 
developed in Ref.\ \cite{KoiNic2023}.

Let us consider a general many-body system which is characterized by 
the canonical pairs of position and momentum in Cartesian coordinates,
$\{q_i,p_i\}_{i=1}^n$.
In the absence of interaction with the environment, the dynamics of this system 
is described by a general Hamiltonian $H$ which can be non-linear. 
Upon introducing this interaction, 
the phase space distribution $\rho(\{q_i,p_i\}_{i=1}^n,t)$ 
is determined by the following 
Fokker-Planck-Kramers (\hypertarget{FPK}{FPK}) equation:
\begin{equation} \label{eq:C_Kramers}
\begin{aligned}
&\partial_t \rho  =  \{ H, \rho \}_{\text{PB}}  +
\sum_{i=1}^n \mathcal{D}_\text{FPK}^{(i)}[\rho] \,, \\
&\mathcal{D}_\text{FPK}^{(i)}[\rho] := \beta_i^{-1}\gamma_{p_i}
\{ {\rm e}^{-\beta_i H}
\{ {\rm e}^{ \beta_i H} \rho, q_i \}_{\text{PB}},\, q_i\}_{\text{PB}} \,\,  +  \\
& \hspace{1.75cm}\beta_i^{-1}\gamma_{q_i}
\{ {\rm e}^{-\beta_i H}
\{ {\rm e}^{\beta_i H} \rho ,  p_i \}_{\text{PB}},  \, p_i \}_{\text{PB}} \, ,
\end{aligned}
\end{equation}
where $\{ \bullet,\bullet\}_{\text{PB}}$ denotes the Poisson bracket and, 
$\gamma_{p_i}$ and $\gamma_{q_i}$ are real parameters.
In this equation,
the $i$-th degree of freedom, described by the pair $( q_i,p_i )$,
interacts with a heat bath with $\beta_i = (k_B T_i)^{-1}$,
where $k_B$ is the Boltzmann constant and $T_i$ is temperature.
When all baths have the same temperature $\beta_1=...=\beta_n=\beta$, 
the stationary state is given by the thermal equilibrium state $\sim {\rm e}^{-\beta H}$
for any Hamiltonian $H$ bounded from below,
including the relativistic Hamiltonian \cite{koide_rbm1,koide_rbm2,deffer}.
Independently of the choice of $\gamma_{q_i}$ and $\gamma_{p_i}$, 
heat, work, and entropy are introduced in such a way that the laws analogous
to the first and second laws of thermodynamics are satisfied \cite{KoiNic2023}.

As shown in Ref.\ \cite{KoiNic2023}, 
the canonical quantization of Eq.\ (\ref{eq:C_Kramers}) 
leads to the quantum master equation
\begin{eqnarray} \label{eq:Q_Kramers}
&& \tfrac{d}{dt}\hat{\rho} (t) =
\tfrac{\ii}{\hbar}\ [ \hat{\rho}(t), \hat{H} ]
+ \sum_{i=1}^n \mathcal{D}^{(i)}_\text{Q}[\hat \rho]\, ,  \\
&& \mathcal{D}^{(i)}_\text{Q}[\hat \rho] :=
- \frac{\gamma_{p_i} }{\beta_i \hbar^2}
[ {\rm e}^{-\tfrac{\beta_i}{2} \! \hat{H}}
[ {\rm e}^{ \tfrac{\beta_i}{2} \! \hat{H}} \, \hat{\rho} \,
  {\rm e}^{ \tfrac{\beta_i}{2} \! \hat{H}}, \hat{q}_i  ]
  {\rm e}^{-\tfrac{\beta_i}{2} \! \hat{H}}, \hat{q}_i  ] \nonumber \\
&& \hspace{1.6cm} - \frac{\gamma_{q_i} }{\beta_i \hbar^2}
[ {\rm e}^{-\tfrac{\beta_i}{2} \! \hat{H}}
[ {\rm e}^{ \tfrac{\beta_i}{2} \! \hat{H}} \, \hat{\rho} \,
  {\rm e}^{ \tfrac{\beta_i}{2} \! \hat{H}}, \hat{p}_i  ]
  {\rm e}^{-\tfrac{\beta_i}{2} \! \hat{H}}, \hat{p}_i  ] \, , \nonumber
\end{eqnarray}
\blue{see also Eq.(\ref{eq:Diss_Q_Kram}).}
When all particles interact with the same heat bath,
or all the baths have the same temperature, 
again the thermal equilibrium state $\sim {\rm e}^{-\beta \hat H}$ 
is a fixed point of this equation.
Moreover, effortlessly, one can show that this 
master equation ensures trace preservation, $\tfrac{d}{dt}{\rm Tr}(\hat \rho) = 0$. 
In the case of harmonic oscillators, 
\begin{equation*}
\hat{H} = \sum_{i=1}^n 
\left(\frac{\hat{p}^2_i}{2m_i} + \frac{m_i\omega^2_i}{2} \hat{q}^2_i \right)\, ,
\end{equation*} 
with $m_i$ being masses and $\omega_i$ angular frequencies,
we have shown that this evolution is \hy{CPTP}, 
by choosing 
$\gamma_{p_i} = (m_i\omega_i)^2 \gamma_{q_i}$, 
{\it i.e.}, it reduces to a \hy{GKSL} form, 
thus ensuring complete positivity \cite{KoiNic2023}.
This condition is reproduced later in Sec.\ \ref{sec:LKE} 
with a systematic procedure.
 
For the sake of later convenience, let us reexpress the \hy{FPK} equation. 
A point in a $2n$-dimensional phase space is denoted as
\begin{equation}\label{eq:x_vec}
x = (q_1,...\,,q_n,p_1,...\,,p_n)^\top \in \mathbb R^{2n} \, ,
\end{equation}
which is the column vector composed by $n$ Cartesian coordinates together
with $n$ canonical conjugate momenta.
The $2n \times 2n$ symplectic matrix is introduced by 
\begin{equation*} %¨\label{eq:mat_J}
{\sf J}:=
\left( \!\! \begin{array}{rc}
       {\bf 0}_{n}   & \mathsf I_{n}  \\
      -\mathsf I_{n} & {\bf 0}_{n}
       \end{array}
\!\! \right), \,\,\, {\sf J}^{-1} = {\sf J}^{\top} = - {\sf J} \, ,
\end{equation*}
where $\mathsf I_{n}$ and ${\bf 0}_{n}$ are, respectively,
the $n \times n$ block identity and block zero matrices.

By opening the Poisson bracket in Eq.\ (\ref{eq:C_Kramers}) 
and  employing the notation from Eq.\  (\ref{eq:x_vec}), 
the \hy{FPK} equation can be succinctly written as the continuity equation.
\begin{equation} \label{eq:C_Kramers2}
\begin{aligned}
&\partial_t \rho (x,t) + \partial_x \! \cdot \! \mathcal{J}_\text{K} = 0,  \\
&\mathcal{J}_\text{K} := 
({\sf I}_{2n} + {\bf C}{\sf J}) ({\sf J} \partial_x H) \rho
               -  {\bf D} \, \partial_x \rho, 
\end{aligned}
\end{equation} 
where $\partial_x := ( \partial_{q_1}, ... ,\partial_{q_n}, 
                 \partial_{p_1}, ... ,\partial_{p_n} )^\top$ 
and
\begin{equation} \label{eq:matrices_CD}
\begin{aligned} 
{\bf C} &:= {\rm Diag}( \gamma_{q_1},...\,, \gamma_{q_n},
                        \gamma_{p_1},...\,, \gamma_{p_n} ) \, ,  \\
{\bf D} &:= 
{\rm Diag}\left( \frac{\gamma_{q_1}}{\beta_1},...\,,\frac{\gamma_{q_n}}{\beta_n},
                 \frac{\gamma_{p_1}}{\beta_1},...\,,\frac{\gamma_{p_n}}{\beta_n} \right)
\, .
\end{aligned}
\end{equation}

A general quadratic Hamiltonian is expressed as
\begin{equation}
H  = \frac{1}{2} (x-\xi) \cdot {\bf H} (x-\xi) + \varphi \, , \label{eqn:quadra_hamil}
\end{equation}
where $\xi \in \mathbb R^{2n}$ is a column vector,
$\varphi \in \mathbb R$ is a constant,
and the Hessian of the Hamiltonian, ${\bf H}_{ij} = \partial^2_{x_i x_j}H(x)$,
is a $2n \times 2n$ symmetric real matrix. 
Using that $\partial_x H = {\bf H} (x-\xi) $ under this setup , 
Eq.\ (\ref{eq:C_Kramers2}) becomes
\begin{align*}
\begin{split}
&\partial_t \rho (x,t) + \partial_x \! \cdot \! \mathcal{J}_\text{K} = 0,  \\
&\mathcal{J}_\text{K} = ( {\sf I}_{2n} + {\bf C}{\sf J} ) {\sf J}{\bf H} (x-\xi)  \rho
               -  {\bf D} \, \partial_x \rho \, .  
\end{split} %\label{eqn:FPK}
\end{align*}

%%%%%%%%%%%%%%%%%%%%%%%%%%%%%%%%%%%%%%%%%%%%%%%%%%%%%%%%%%%%%%%%%%%%%%%%%%%%%%%%%%%%%%%%%
%%%%%%%%%%%%%%%%%%%%%%%%%%%%%%%%%%%%%%%%%%%%%%%%%%%%%%%%%%%%%%%%%%%%%%%%%%%%%%%%%%%%%%%%%
\section{Classical Limit of the Quantum Evolution} \label{sec:CLotQE}         %%%%%%%%%%%
%%%%%%%%%%%%%%%%%%%%%%%%%%%%%%%%%%%%%%%%%%%%%%%%%%%%%%%%%%%%%%%%%%%%%%%%%%%%%%%%%%%%%%%%%
%%%%%%%%%%%%%%%%%%%%%%%%%%%%%%%%%%%%%%%%%%%%%%%%%%%%%%%%%%%%%%%%%%%%%%%%%%%%%%%%%%%%%%%%%
Before discussing complete positivity, 
let us first confirm the classical limit 
of our quantum master equation (\ref{eq:Q_Kramers}) 
\blue{for a general Hamiltonian}.

Canonical quantization promotes the column vector (\ref{eq:x_vec}) 
to the following column vector of operators:
\begin{equation}\label{eq:vec_op_x}
\hat x := (\hat q_1,...\,,\hat q_n, \hat p_1,...\hat p_n)^\top \, ,
\end{equation}
and each of its $2n$ components, generically called $\hat x_j $, is Hermitian.
In this notation, the canonical commutation relations
are compactly written as
\begin{equation}\label{eq:ComRel}
[\hat x_j , \hat x_k ] = \ii \hbar \, \mathsf J_{jk} \, .
\end{equation}

Expanding the exponentials in Eq.\ (\ref{eq:Q_Kramers}), 
retaining the terms up to first order in $\beta_i$, 
and employing the column vector in Eq.\ (\ref{eq:vec_op_x}), 
we find the quantum master equation in high temperature limit,
\begin{equation}\label{eq:H_Temp_Q_Kram}
\begin{aligned}
& \tfrac{d}{dt}\hat{\rho} (t) =
\tfrac{\ii}{\hbar}\ [ \hat{\rho}(t), \hat{H} ]
+ \sum_{i=1}^n \mathcal{D}^{(i)}_0[\hat \rho] + 
  \sum_{i=1}^n \mathcal{D}^{(i)}_1[\hat \rho] \, ,   \\
& \mathcal{D}^{(i)}_0[\hat \rho]  := 
 - \frac{1}{{\hbar}}\sum_{j=1}^{2n} 
   {\bf K}_{ij} \left[ \{ \hat \rho, \hat x_i \hat x_j \} 
                       - 2\hat x_i \hat \rho \hat x_j    \right] \, ,   \\     
& \mathcal{D}^{(i)}_{1}[\hat \rho]  :=  
\frac{ \beta_i}{2\hbar} \sum_{j=1}^{2n}{\bf K}_{ij} 
                 \left[ \{ \hat \rho, [\hat x_i,\hat H]\}, \hat x_j \right] \, , 
%
%&& +  \frac{1}{{2\hbar}} \sum_{j=1}^{2n}   
%{\bf K}_{ij} \beta_j \hat x_i \{ \hat \rho, [\hat H, \hat x_j]\} \, ,  
\end{aligned}
\end{equation}
where $\{ \bullet,\bullet\}$ denotes the anti-commutator and
\begin{equation} \label{eq:mat_K}
{\bf K} = \frac{1}{\hbar}{\sf J}{\bf D}{\sf J}^\top \! 
= {\rm Diag} \! \left( \frac{\gamma_{p_1}}{\hbar\beta_1}, ... ,
                       \frac{\gamma_{p_n}}{\hbar\beta_n},
                       \frac{\gamma_{q_1}}{\hbar\beta_1}, ... ,
                       \frac{\gamma_{q_n}}{\hbar\beta_n}        \right) .
\end{equation}
Noteworthy, 
the zeroth order term $\mathcal{D}^{(i)}_0[\hat \rho]$ 
is a generator of a \hy{GKSL} equation for the Lindblad operator
$\hat L_i = \sqrt{ 2 {\bf K}_{ii} } \, \hat x_i$.

Considering the Wigner function $W(x,t)$ for the density operator $\hat \rho(t)$,  
the Wigner symbol of each term in Eq.\ (\ref{eq:H_Temp_Q_Kram}) 
may be determined using the recipe in Appendix \ref{app:WWC}. 
Note that the Wigner symbol of the operator $\hat x$ 
is the column vector $x$ in Eq.\ (\ref{eq:x_vec}) \cite{ozorio1998} 
and $H(x)$ is the symbol of the Hamiltonian operator $\hat H$.
Then the products of operators in Eq.\ (\ref{eq:H_Temp_Q_Kram}) are 
evaluated using the Groenewold formula as
\begin{equation*}
\begin{aligned}
& \tfrac{d}{dt} \hat \rho (t) \rightarrow \partial_t W(x,t)  \, ,  \\ 
&\tfrac{\ii}{\hbar}\ [ \hat\rho(t), \hat H ] \rightarrow 
- \{ W(x,t), H(x)\}_\text{PB} + O(\hbar^2) \\
& \{ \hat \rho, [\hat H, \hat x_i] \} \hat x_j \rightarrow 
-{2i\hbar} W(x,t) \dot x_i x_j + 
\hbar^2 ({\sf J} \partial_{x_j} W ) \dot x_i \\
& \qquad \qquad \qquad + \frac{\hbar^2}{2}W(x,t) 
          [{\sf J} {\bf H}(x) {\sf J}^\top]_{ij} + O(\hbar^2) \, ,       \\
& \hat x_i\{ \hat \rho, [\hat H, \hat x_j] \}  \rightarrow 
-2i\hbar W(x,t) x_i \dot x_j   
- \hbar^2 ({\sf J} \partial_{x_i} W ) \dot x_j \\
& \qquad \qquad \qquad - \frac{\hbar^2}{2}W(x,t)  
           [{\sf J} {\bf H}(x) {\sf J}^\top]_{ji} + O(\hbar^2) \, ,   \\   
&\{ \hat \rho, \hat x_i \hat x_j \} 
- 2\hat x_i \hat \rho \hat x_j   \rightarrow 
\hbar ({\sf J} \partial_{x})_i ({\sf J} \partial_{x})_j W(x,t) \, ,
\end{aligned}
\end{equation*}
where $\{ W, H\}_\text{PB} := \partial_x W \cdot {\sf J} \partial_x H$ \cite{ozorio1998}
and 
${\bf H}(x) := \partial_{xx} H(x)$ is the Hessian of the symbol $H(x)$, 
a classical Hamiltonian.
See the discussion Appendix \ref{app:WWC} for details.
The first line in the above list represents an exact expression, 
as it directly corresponds to the mapping of the density operator 
to its corresponding Wigner function $W(x,t)$. 
The last line is also an exact representation, but for a different reason: 
in this case, terms of order $\hbar^2$ and higher vanish 
in the Groenewold product formula.
All other lines are semiclassical expansions, 
through the Groenewold formula, 
keeping terms of $O(\hbar)$.

In the end, by collecting all the symbols mentioned above and 
using ${\bf K}$ from Eq.\ (\ref{eq:mat_K}), 
the Wigner transformation of Eq.\ (\ref{eq:H_Temp_Q_Kram}), 
while discarding $O(\hbar^2)$ terms, becomes 
\begin{equation}\label{eq:classlim}
\begin{aligned}
&\partial_t W(x,t) + \partial_x \! \cdot \! {\mathcal J}_W = 0 \, , \\
&{\mathcal J}_W := ( {\sf I}_{2n} + {\bf C} {\sf J})[{\sf J} \partial_x H(x)]\, W 
                   - {\bf D}\, \partial_x W \, .
\end{aligned}
\end{equation}
One can see that the above equation for the Wigner function matches 
the FPK equation (\ref{eq:C_Kramers2}). 
When all the temperatures of the baths are the same, 
the FPK equation (\ref{eq:C_Kramers}) describes thermal relaxation processes, 
as shown in Ref.\ \cite{KoiNic2023}.
Due to this equivalence, the same thermal relaxation 
is observed in Eq.\ (\ref{eq:classlim}). 
In other words, the classical limit of 
the quantum master equation (\ref{eq:Q_Kramers}) describes 
the same thermal relaxation process as 
the FPK equation (\ref{eq:C_Kramers}), 
despite the issue that the Wigner function 
does not satisfy the positive semidefinite property 
of true probability densities. 

In the discussion so far, we have assumed that the matrix ${\bf C}$ 
in Eq.\ (\ref{eq:matrices_CD}) does not depend on the bath temperatures.
However it is also possible to consider temperature 
dependence as discussed in Sec.\ref{sec:EC}. 
In the context of this section, the additional temperature dependences, 
which come from Eq.\ (\ref{eqn:add-tempdep_gamma}), 
affect only the higher order terms beyond $O(\hbar^2)$ 
and thus the above conclusion is not affected.

%%%%%%%%%%%%%%%%%%%%%%%%%%%%%%%%%%%%%%%%%%%%%%%%%%%%%%%%%%%%%%%%%%%%%%%%%%%%%%%%%%%%%%%%%
%%%%%%%%%%%%%%%%%%%%%%%%%%%%%%%%%%%%%%%%%%%%%%%%%%%%%%%%%%%%%%%%%%%%%%%%%%%%%%%%%%%%%%%%%
\section{CPTP Criterion}        \label{sec:GKSLME}    %%%%%%
%%%%%%%%%%%%%%%%%%%%%%%%%%%%%%%%%%%%%%%%%%%%%%%%%%%%%%%%%%%%%%%%%%%%%%%%%%%%%%%%%%%%%%%%%
%%%%%%%%%%%%%%%%%%%%%%%%%%%%%%%%%%%%%%%%%%%%%%%%%%%%%%%%%%%%%%%%%%%%%%%%%%%%%%%%%%%%%%%%%

A general quadratic Hamiltonian operator corresponding 
to Eq.\ (\ref{eqn:quadra_hamil}) is given by 
\begin{equation} \label{eq:quad_Ham}
\hat H =
\frac{1}{2} (\hat x-\xi) \cdot {\bf H}  (\hat x-\xi) + \varphi\hat 1 \, .
\end{equation}
See the related discussion in Appendix \ref{app:QH}.
With the quantum master equation (\ref{eq:Q_Kramers}) in mind for such Hamiltonian,   
we will consider the 
quadratic time-convolutionless \hypertarget{QTCL}{(QTCL)} master equation, 
\begin{equation} \label{eq:QTCL}
\begin{aligned}
&\tfrac{d}{dt}\hat{\rho} (t) = \tfrac{\ii}{\hbar}\ [ \hat{\rho}(t), \hat{H} ]
                             + \mathcal{D}[\hat \rho] \, ,  \\
&\mathcal{D}[\hat \rho] = 
- \frac{1}{\hbar} \sum_{j,k=1}^{2n} {\bf \Xi}_{jk}^\top
                   \hat \rho (\hat x_j - \eta_j )
                             (\hat x_k - \eta_k ) \\
& \hspace{1.cm}+ \frac{1}{\hbar} \sum_{j,k=1}^{2n} ({\bf \Xi} + {\bf \Xi}^\dag)_{jk}
                    (\hat x_j - \eta_j) \hat \rho
                    (\hat x_k - \eta_k) \\
& \hspace{1.cm}- \frac{1}{\hbar} \sum_{j,k=1}^{2n} {\bf \Xi}_{jk}^\ast
                    (\hat x_j - \eta_j)
                    (\hat x_k - \eta_k) \hat \rho  \, ,
\end{aligned}
\end{equation}
where ${\bf \Xi}_{jk}$ are the elements of a $2n \times 2n$ 
complex matrix $\bf \Xi$ and $\eta \in \mathbb R^{2n}$, 
which can be time-dependent due to the time dependences 
of the parameters in the Hamiltonian (\ref{eq:quad_Ham}).
See Appendix \ref{app:QuadGen} for details.
Our quantum master equation (\ref{eq:Q_Kramers}) 
is reproduced by choosing this matrix and column vector appropriately.
%\red{Then ${\bf \Xi}_{jk}$ can be a function of time. }
For the moment, however, we will continue the discussion without their specifications. 

Performing the decomposition
\begin{equation}\label{eq:mat_Xi-h-a}
{\bf \Xi} = \tfrac{1}{2}({\bf \Xi}_\text{H} + {\bf \Xi}_\text{A}) \, , 
\end{equation}
where 
\[
\begin{aligned}
{\bf \Xi}_\text{H} := {\bf \Xi} + {\bf \Xi}^{\dag} \,, \\
{\bf \Xi}_\text{A} := {\bf \Xi} - {\bf \Xi}^{\dag} \,, 
\end{aligned}
\]
the \hy{QTCL} equation is written as
\begin{equation}\label{eq:Q_Quad_Kram1}
\begin{aligned}
\tfrac{d}{dt}\hat{\rho}
&=
\tfrac{\ii}{\hbar} [ \hat{\rho} \, , \hat{H}_\text{eff} ]  
- \tfrac{1}{2\hbar} \! \left\{ \hat \rho \, , (\hat x -\eta)\cdot
{\bf \Xi}_\text{H}^\ast  (\hat x -\eta ) \right\}                      \\
&+ \tfrac{1}{\hbar} \sum_{i,j =1}^{2n}
({\bf \Xi}_\text{H})_{ij} (\hat x_i -\eta_i)  \hat \rho
                        (\hat x_j -\eta_j)    \, ,
\end{aligned}
\end{equation}
where 
\begin{equation}\label{eq:Heff}
\hat{H}_\text{eff} := \hat{H} - 
\tfrac{\ii}{2}(\hat x -\eta) \cdot {\bf \Xi}_\text{A}^\ast (\hat x -\eta ) \, .
\end{equation}
Note that $\ii \,{\bf \Xi}_\text{A}^\ast$ is Hermitian and that 
such shift of the system Hamiltonian, 
induced by the interaction with the environment, 
is commonly observed in various different approaches \cite{dieter,BreuerPetr2002}.

Due to the Hermiticity of ${\bf \Xi}_\text{H}$, 
one can introduce the eigenvectors ${v}_\mu \in \mathbb C^{2n}$ as 
\begin{equation*}
{\bf \Xi}_\text{H} {v}_\mu = a_\mu {v}_\mu \, , \,\,\, \mu =1,...,2n \, ,
\end{equation*}
where $a_\mu \in \mathbb R$ are the respective eigenvalues 
and the matrix ${\bf \Xi}_\text{H}$ can be represented by 
\begin{equation}\label{eq:mat_Xih}
{\bf \Xi}_\text{H} = \sum_{\mu=1}^{2n} a_\mu {v}_\mu {v}_\mu^\dag =
\sum_{\mu = 1}^{2n} \lambda_\mu g_{\mu\nu} \lambda_\nu^\dag\, ,
\end{equation}
where $g_{\mu\nu} := \text{sgn}(a_\mu) \delta_{\mu\nu}$.
In short, the master equation in Eq.\ (\ref{eq:Q_Quad_Kram1})
takes the form of a general time-convolutionless 
\hypertarget{GTCL}{(GTCL)} equation \cite{breuer2016}:
\begin{align}\label{eq:GTCL}
\tfrac{d}{dt}\hat{\rho} =
\tfrac{\ii}{\hbar} [ \hat{\rho}, \hat{H}_{\text{eff}} ]
- \! \sum_{\mu,\nu =1}^{2n}\frac{g_{\mu\nu}}{2\hbar}
\left( \{ \hat \rho, \hat L_\mu^\dag \hat L_\nu \}
       -2  \hat L_\mu \hat \rho \hat L_\nu^\dag \right) \, ,
\end{align}
where 
$\hat L_\mu = \lambda_\mu \cdot (\hat x - \eta )$ 
are the Lindblad operators representing the action of the 
environment in the system evolution.

Now we are in a position to examine 
the criterion on the parameters in the \hy{QTCL} equation  (\ref{eq:Q_Quad_Kram1}) 
[or equivalently, Eq.\ (\ref{eq:GTCL})] to ensure that 
the resulting time evolution is a \hy{CPTP} map.
As well-known \cite{breuer2016,rivas2012}, 
a \hy{GKSL} equation is given by a \hy{GTCL} equation (\ref{eq:GTCL}) 
for $g_{\mu\nu} = \delta_{\mu\nu}$. 
To satisfy this, 
we should require that $a_\mu \ge 0, \forall \mu$ in Eq.\ (\ref{eq:mat_Xih}).
Consequently, the time evolution of the \hy{QTCL} equation 
is a \hy{CPTP} map if and only if
\begin{equation}\label{eq:CondGKSL}
{\bf \Xi}_\text{H} \ge 0 \, , 
\end{equation}
which means that its eigenvalues are non-negative.

For a given evolution of $\hat \rho$, 
the definitions of $\hat H$ and ${\cal D}[\hat{\rho}]$ in Eq.\ (\ref{eq:QTCL}),   
or equivalently the definitions of $\hat H_\text{eff}$ 
and $\hat L_\mu$ in Eq.\ (\ref{eq:GTCL}),  
are known to be non-unique \cite{colla}.
Consequently, the definition of ${\bf \Xi}_\text{H}$ may be ambiguous. 
However, as we show in Appendix \ref{app:non-unique}, 
the matrix ${\bf \Xi}_\text{H}$ 
remains invariant despite the aforementioned nonuniqueness, 
thereby ensuring the validity of the derived condition for a general \hy{QTCL} equation.

When the \hy{GTCL} equation is given by our quantum master equation (\ref{eq:Q_Kramers})
with the Hamiltonian (\ref{eq:quad_Ham}),
we find $\eta = \xi$ and the matrix ${\bf \Xi}$ defined by 
\begin{align}\label{eq:mat_Xi}
{\bf \Xi}_{ij} 
&= \sum_{k=1}^{2n} {\bf K}_{ik}({\sf S}_{\!\beta_k})_{kj}
             =  {\bf K}_{ii} ({\sf S}_{\!\beta_i})_{ij} \, ,
\,\,\,
(\beta_{i+n} = \beta_n) \, ,
\end{align}
where $\bf K$ is defined in Eq.\ (\ref{eq:mat_K}) and 
\begin{equation*} %\label{eq:mat_S}
\begin{aligned}
{\sf S}_{\!\beta_k} &:= \exp[{\ii \hbar \beta_k } {\sf J}{\bf H}/2] \, ;
\end{aligned}
\end{equation*}
see Appendix \ref{app:QH}. 
Note that the different rows of ${\bf \Xi}$ depend on different values
of temperature, {\it e.g.},
\begin{align}
\begin{split}
{\bf \Xi}_{12} &= {\bf K}_{11} ({\sf S}_{\!\beta_1})_{12}
             = \frac{\gamma_{p_1}}{\hbar\beta_1} ({\sf S}_{\!\beta_1})_{12} \, , \\
{\bf \Xi}_{21} &= {\bf K}_{22} ({\sf S}_{\!\beta_2})_{12}
             = \frac{\gamma_{p_2}}{\hbar\beta_2} ({\sf S}_{\!\beta_2})_{21} \, ,
\end{split}\nonumber
\end{align}
and 
\begin{align}
{\bf \Xi}^\dagger_{ij} 
= {\bf \Xi}^\ast_{ji} 
&= {\bf K}_{jj} ({\sf S}_{\!\beta_j}^\ast)_{ji} =
{\bf K}_{jj} ({\exp}[\ii \hbar \beta_j {{\sf J}{\bf H} }/2])_{ji} \nonumber \\
& = {\bf K}_{jj} ({\exp}[-\ii \hbar \beta_j {{\bf H}{\sf J}}/2])_{ij} \, \nonumber.
\end{align}

The criterion (\ref{eq:CondGKSL}) imposes a constraint on 
the system-bath interaction parameters, as detailed in Eq.\ (\ref{eq:mat_K}). 
We can assert that the bath-momentum couplings $\gamma_{p_i}$ 
are intrinsically linked to the position-bath couplings $\gamma_{q_i}$, 
as anticipated in Sec.\ref{sec:CQoGBM} for the harmonic oscillator.
In the end, 
the criterion determines the permissible values for a genuine physical evolution 
represented by a \hy{CPTP} map.

To study the meaning of above criterion in the corresponding classical theory,
let us remember the discussion in Sec.\ \ref{sec:CLotQE}. 
As discussed in Ref.\ \cite{KoiNic2023}, the \hy{FPK} equation permits any value of
$\gamma_{q_i}$ and $\gamma_{p_i}$, differently from its quantum counterpart.
Therefore, the criterion (\ref{eq:CondGKSL}) will be automatically satisfied
in the classical limit of the matrix ${\bf \Xi}$.
Expanding the matrix ${\mathsf S}_{\!\beta_k}$ in Eq.\ (\ref{eq:mat_Xi}), 
${\bf \Xi}_\text{H}$ is given by 
\begin{equation*}
{\bf \Xi}_{\rm H} = {\bf K} > 0 \, ,
\end{equation*}
where we used 
\begin{equation*}
{\bf \Xi} \approx {\bf K} -\tfrac{\ii}{2} {\sf J}  {\bf C}{\bf H} \, .
\end{equation*}
It is easy to see that the criterion (\ref{eq:CondGKSL}) 
is satisfied in the classical limit.
Again, we can employ the analogy with the uncertainty principle 
that does not provide constraints to classical variables.

Our master equation (\ref{eq:Q_Kramers}) 
drives the system towards thermal equilibrium 
when $\beta_1 = \beta_2 = ... = \beta_n = \beta$, 
at least in its classical limit. 
However, an additional condition is required to ensure complete positivity.
In standard approaches, the situation is inverted: 
one assumes a \hy{CPTP} equation and imposes additional constrains, 
called quantum detailed balance conditions \cite{QDBCfin},  
to describe a thermal relaxation process. 
In Appendix \ref{app:QDBC}, as a consequence of detailed balance, 
we show that a \hy{CPTP} evolution driven by our quantum master equation 
for the quadratic Hamiltonian (\ref{eq:quad_Ham}) 
will attain thermal equilibrium if this Hamiltonian has 
a positive definite Hessian, ${\bf H} > 0$, 
which includes the cases of harmonic oscillators 
and linearly coupled oscillators called the network model. 

%%%%%%%%%%%%%%%%%%%%%%%%%%%%%%%%%%%%%%%%%%%%%%%%%%%%%%%%%%%%%%%%%%%%%%%%%%%%%%%%%%%%%%%%%
%%%%%%%%%%%%%%%%%%%%%%%%%%%%%%%%%%%%%%%%%%%%%%%%%%%%%%%%%%%%%%%%%%%%%%%%%%%%%%%%%%%%%%%%%
\section{Bloch-Redfield Equation} \label{sec:AtRE}
%%%%%%%%%%%%%%%%%%%%%%%%%%%%%%%%%%%%%%%%%%%%%%%%%%%%%%%%%%%%%%%%%%%%%%%%%%%%%%%%%%%%%%%%%
%%%%%%%%%%%%%%%%%%%%%%%%%%%%%%%%%%%%%%%%%%%%%%%%%%%%%%%%%%%%%%%%%%%%%%%%%%%%%%%%%%%%%%%%%
The Bloch-Redfield equation is derived from a microscopic system-bath Hamiltonian 
using Born-Markov approximations \cite{BreuerPetr2002}, 
assuming that the bath degrees of freedom are thermally equilibrated.
Further approximations often lead to a \hy{CPTP}  equation. 
For a standard discussion and various improvements, see Refs.
\cite{BreuerPetr2002,abbruzzo,becker,davidovic,pradilla,breuer04,gneiting,piilo,whitney}.

In contrast, 
the derivation of our quantum master equation (\ref{eq:Q_Kramers}) 
is more phenomenological, since it is obtained by applying a canonical quantization 
to the classical model of a generalized Brownian motion (\ref{eq:C_Kramers}), 
which results from the coarse-graining of classical degrees of freedom.

Considering Hermitian system operators $\hat{A}_{i}$ ($i=1,2,\cdots$),
the Bloch-Redfield equation is defined by
\begin{equation}\label{eqn:redfield}
\begin{aligned} 
& \partial_t \hat \rho (t ) =
\frac{\ii}{\hbar}  [ \hat \rho (t), \hat H]  + {\cal R}[\hat{\rho}] \, , \\
&{\cal R}[\hat{\rho}]  =  
-\frac{1}{\hbar^2} \sum_{i,j}\int^\infty_0 \!\! d\tau \, {\bf \Gamma}_{ij} (\tau) 
[\hat A_{i}, \tilde A_{j}(t) \hat \rho  ] + \text{h.c.} \, ,
\end{aligned}
\end{equation}
where the functions 
${\bf \Gamma}_{ij} (s)$ are characterized by the temporal correlation functions 
of the bath degrees of freedom and 
$\tilde A_{j}(t) := 
{\rm e}^{\frac{\ii}{\hbar} \hat{H} t} \hat A_{j} 
{\rm e}^{-\frac{\ii}{\hbar} \hat{H} t}$. 
In this form, 
the Markov approximation has already been employed, considering that 
the time scale of the system evolution is much slower than the one of 
the bath degrees of freedom. 

The dissipative terms in Eq.\ (\ref{eq:Q_Kramers}) can be rewritten as 
\begin{equation*}
{\mathcal D}_\text{Q}^{(i)}[\hat{\rho}] =
-\frac{1}{\hbar} \sum_{j=1}^{2n}{\bf K}_{ij} 
[ \hat x_i, {\rm e}^{-\frac{\beta_j}{2} \hat H} \hat x_j 
            {\rm e}^{\frac{\beta_j}{2} \hat H} \hat \rho] + \text{h.c.}\, .
\end{equation*}
In Appendix \ref{app:QuadGen}, 
it is shown that the quantum master equation (\ref{eq:Q_Kramers}) and 
the Bloch-Redfield equation have a very similar structure 
if the system operators in the Bloch-Redfield equation are chosen to be  
linear operators in $\hat{x}$, {\it i.e.}, 
$\hat{A}_{i} = a_{i} \hat x_i $, for a real vector $a \in \mathbb R^{2n}$. 
However, it is important to note that 
${\cal D}^{(i)}_\text{Q} [\hat{\rho}]$ 
does not exhibit any oscillatory behavior induced by $\hat{H}$. 
In the standard approach \cite{BreuerPetr2002}, 
a \hy{CPTP} evolution is achieved from the Bloch-Redfield equation by applying 
the rotating wave approximation, which eliminates non-secular terms associated 
with oscillations caused by ${\rm e}^{\pm\frac{\ii}{\hbar} \hat{H} t}$. 

Let us apply our \hy{CPTP} criterion (\ref{eq:CondGKSL}) 
to the Bloch-Redfield equation (\ref{eqn:redfield}).
For a quadratic Hamiltonian and linear operators $\hat{A}_{i}$, 
it is possible to write ${\cal R}[\hat{\rho}]$ above as ${\cal D}[\hat{\rho}]$ 
in \hy{QTCL} (\ref{eq:QTCL}) for the matrix ${\bf \Xi}$ with elements 
\[ 
{\bf \Xi}_{ij} = \frac{1}{\hbar}\sum_{k=1}^{2n}\int_{0}^\infty \!\! d\tau \, 
 a_i {\bf \Gamma}_{ik}^\ast(\tau) a_k \,  ({\sf S}_t)_{kj} \, ,  
\]
where $\tilde A_{j}(t)$ and ${\sf S}_t$ are represented using 
Eqs.\ (\ref{eq:CovRules}) and (\ref{eq:Symp_S}). 
Consequently, to satisfy the criterion (\ref{eq:CondGKSL}), 
the column vector $a$ should be chosen appropriately for a given matrix ${\bf \Gamma}$.

Although no one has yet succeeded in finding such parameters, 
there have been several attempts to minimize the violation of the \hy{CPTP} map
by selecting appropriate parameters.
The generator ${\cal D}[\hat{\rho}]$ in Eq.\ (\ref{eq:QTCL}) 
can be expanded using variables differently from $(\hat{x}_{i} - \eta_i)$. 
In Ref.\ \cite{becker}, 
the expansion in terms of $\hat A_{i} + \alpha_i(t) \tilde A_{i}$ 
is considered and  
a non-\hy{CPTP} time-convolutionless master equation is obtained. 
A \hy{CPTP} equation is derived from this by 
truncating the terms that violate positivity using 
the parameters $\alpha_i(t)$ which minimize the contributions from 
these non-positivity terms.
%The derived equation is mapped into Eq.\ (\ref{eq:GTCL}) with $g_{\mu\mu} = -1$.
A similar approach is considered in Ref. \ \cite{abbruzzo}, 
which uses a different expansion and discusses the relation 
to the result in Ref.\ \cite{becker}.
Other related approaches and decompositions are performed in  
Refs. \cite{whitney,davidovic,pradilla,breuer04,gneiting,piilo}. 

\

%%%%%%%%%%%%%%%%%%%%%%%%%%%%%%%%%%%%%%%%%%%%%%%%%%%%%%%%%%%%%%%%%%%%%%%%%%%%%%%%%%%%%%%%%
%%%%%%%%%%%%%%%%%%%%%%%%%%%%%%%%%%%%%%%%%%%%%%%%%%%%%%%%%%%%%%%%%%%%%%%%%%%%%%%%%%%%%%%%%
\section{Applications to $n=1$} \label{sec:LKE} %%%%%%%%%%%%%%%%%%%%%%%%%%%%%%%%%%%%%%%%%
%%%%%%%%%%%%%%%%%%%%%%%%%%%%%%%%%%%%%%%%%%%%%%%%%%%%%%%%%%%%%%%%%%%%%%%%%%%%%%%%%%%%%%%%%
%%%%%%%%%%%%%%%%%%%%%%%%%%%%%%%%%%%%%%%%%%%%%%%%%%%%%%%%%%%%%%%%%%%%%%%%%%%%%%%%%%%%%%%%%
Let us apply our criterion to a quadratic Hamiltonian of a system with 
one-degree of freedom ($n=1$). We then  find 
\begin{align*}
\begin{split}
{\mathsf S}_\beta &= \cosh(\Theta) {\sf I}_{2}  
-\ii \frac{\sinh\Theta}{\Theta} {\sf J} {\bf H} \, , \\
{\bf \Xi}_\text{H}& =
2\cosh(\Theta) \, {\bf K}
-  \ii (\gamma_q H_{11} + \gamma_p H_{22}) \frac{\sinh\Theta}{ 2\Theta} {\sf J} \, , \\
{\bf \Xi}_\text{A} &=  \ii \,
\frac{\sinh\Theta}{ 2\Theta}
                \begin{pmatrix}
                -2\gamma_p H_{12} & \gamma_q H_{11} - \gamma_p H_{22} \\
                 \gamma_q H_{11} - \gamma_p H_{22} & 2\gamma_q H_{12}
                \end{pmatrix} \, ,
\end{split}
\end{align*}
where $H_{ij}$ denote the elements of the Hessian $\bf H$ and 
\begin{equation}
\Theta = \frac{\hbar \beta}{2} \sqrt{\det{\bf H}} \, . \label{eq:mat_Xi-h-a_1d}
\end{equation}
The indeces $i=1$ in $\beta$, $\gamma_p$ and $\gamma_q$ are omitted for simplicity.
In this calculation, we used  $({\sf J} {\bf H})^2 = - (\det{\bf H}) {\sf I}_{2}$ 
for a $2 \times 2$ matrix $\bf H$.

%%%%%%%%%%%%%%%%%%%%%%%%%%%%%%%%%%%%%%%%%%%%%%%%%%%%%%%%%%%%%%%%%%%%%%%%%%%%%%%%%%%%%%%%%
\subsection{Elliptic case: $\det{\bf H} >0$}\label{sec:EC}
%%%%%%%%%%%%%%%%%%%%%%%%%%%%%%%%%%%%%%%%%%%%%%%%%%%%%%%%%%%%%%%%%%%%%%%%%%%%%%%%%%%%%%%%%
We first consider the case of $\det{\bf H} >0$.
%It is then easy to see that ${\bf \Xi}_\text{A}$ is pure imaginary and thus
%the second condition in Eq.\ (\ref{eq:CondGKSL}) is satisfied.
The eigenvalues of ${\bf \Xi}_\text{H}$ are given by 
\begin{align*}
 \psi_{\pm}({\bf \Xi}_\text{H})  
= \tfrac{1}{2}\tr{\bf \Xi}_\text{H} \pm
  \tfrac{1}{2}\sqrt{(\tr{\bf \Xi}_\text{H})^2 -4 \det{{\bf \Xi}_\text{H}}}  \, ,
\end{align*}
where
\begin{align}
\begin{split}
\tr{\bf \Xi}_\text{H} &= 
\frac{2\cosh\Theta}{\hbar \beta}  \, {\rm Tr}{\bf C} \ge 0 \, , \\
\det{{\bf \Xi}_\text{H}} &= 
\frac{4\cosh^2\!\Theta}{\hbar^2\beta^2}  \det{\bf C}
- \left[\frac{\sinh \Theta}{2\Theta } \, {\rm Tr}({\bf C}{\bf H}) \right]^2\, ,
\end{split} \nonumber
\end{align}
where ${\bf C}$ is defined by Eq.\ (\ref{eq:matrices_CD}). 

Note that ${\bf \Xi}_\text{H}$ is Hermitian and thus its eigenvalues are real, 
$(\tr{\bf \Xi}_\text{H})^2 \ge 4 \det{{\bf \Xi}_\text{H}}$. 
Therefore, to satisfy our criterion (\ref{eq:CondGKSL}), we require 
\begin{equation*}
\det{{\bf \Xi}_\text{H}} \ge 0 \Longleftrightarrow
\frac{1}{4} \frac{(\tr{{\bf CH}})^2}{\det{{\bf CH}}} \tanh^2 \Theta \le 1 \, ,
\end{equation*}
which for an arbitrary temperature becomes 
\begin{equation}\label{eqn:cond_onedegree}
\frac{1}{4} \frac{(\tr{{\bf CH}})^2}{\det{{\bf CH}}} \le 1 \, ,
\end{equation}
since $\tanh \Theta \le 1$. 

For example, when the Hamiltonian satisfies $H_{12}=0$,
noting that $H_{21}=H_{12}$, 
Eq.\ (\ref{eqn:cond_onedegree}) leads to 
\begin{equation*}
\frac{\gamma_q H_{11}}{\gamma_p H_{22}} + 
\frac{\gamma_p H_{22}}{\gamma_q H_{11}} \le 2 \, .
\end{equation*}
Therefore,  
the quantum master equation (\ref{eq:Q_Kramers}) becomes a \hy{CPTP} evolution for any temperature of heat bath 
when $\gamma_q H_{11} = \gamma_p H_{22}$. 
For a harmonic oscillator, 
where $H_{11} = m\omega^2$ and $H_{22}=1/m$,  
this condition reproduces the result in Ref.\ \cite{KoiNic2023},
\begin{equation}
\gamma_p = (m\omega)^2 \gamma_q \, . \label{eqn:cond-originalho}
\end{equation}

Even after employing the condition (\ref{eqn:cond-originalho}), 
there still remains some freedom for the choice of the parameters 
which can be utilized to introduce further temperature dependence.
For example, let us assume 
\begin{equation}
\gamma_p = \frac{m \beta \hbar \omega}
                {2\sinh( \beta \hbar \omega/2)}  
 \tilde{\gamma} \, ,
\label{eqn:add-tempdep_gamma}
\end{equation}
where $\tilde{\gamma}$ is a non-negative constant. 
Using Eqs.\ (\ref{eqn:cond-originalho}) and (\ref{eqn:add-tempdep_gamma}), 
the quantum master equation (\ref{eq:Q_Kramers})
for a harmonic oscillator becomes
\begin{align} \label{eq:Optics_ME}
\tfrac{d}{dt}\hat{\rho} 
&=  \frac{\ii}{\hbar} [\hat{\rho},\hat{H}]
-
\ii \frac{\tilde{\gamma}}{4\hbar}
\left( [\hat{q}, \{\hat{p}, \hat{\rho}\} ] - [\hat{p}, \{\hat{q}, \hat{\rho}\} ]\right)
\nonumber \\
&  
- \frac{\tilde{\gamma}}{4\hbar} \coth( \tfrac{1}{2}\beta \hbar \omega ) 
\left( 
m\omega [\hat{q}, [\hat{q},\hat{\rho}]] +
\frac{1}{m\omega} [\hat{p}, [\hat{p},\hat{\rho}]]
\right) 
 \, ,
\end{align}
which is essentially the same as the master equation for a harmonic oscillator
interacting with thermal radiation through
the Born-Markov approximations \cite{BreuerPetr2002,ScullyZubairy}.
Indeed, this is rewritten as Eq.\ (\ref{eq:GTCL})
by choosing $g_{\mu\nu} = \delta_{\mu\nu}$, $\eta = 0$, and
\begin{align}\label{eq:OHLind}
\begin{split}
\hat L_1 
&= \sqrt{\hbar \tilde{\gamma} (\bar n + 1)} \, \hat a \, , \\
\hat L_2
&= \sqrt{\hbar \tilde{\gamma} \bar n } \, \hat a^\dagger  
\end{split}
\end{align}
with 
$\bar n = [\exp(\hbar \beta \omega)-1]^{-1}$ and 
$\hat a =  \sqrt{m \omega/(2\hbar)} \hat q +
           \ii \hat p/ \sqrt{2\hbar m\omega}$.
Noteworthy,  
\begin{equation*}
\lim_{\hbar\rightarrow 0} \gamma_p = m \tilde{\gamma} 
\end{equation*}
and thus the classical limit of the master equation (\ref{eq:Optics_ME}) works 
as well as the discussion in Sec.\ \ref{sec:GKSLME}.

%%%%%%%%%%%%%%%%%%%%%%%%%%%%%%%%%%%%%%%%%%%%%%%%%%%%%%%%%%%%%%%%%%%%%%%%%%%%%%%%%%%%%%%%%
\subsection{Hyperbolic and parabolic cases: $\det{\bf H} \le 0$}
%%%%%%%%%%%%%%%%%%%%%%%%%%%%%%%%%%%%%%%%%%%%%%%%%%%%%%%%%%%%%%%%%%%%%%%%%%%%%%%%%%%%%%%%%
For the hyperbolic case, $\det{\bf H} <0$,
we replace $\sqrt{\det{\bf H}}$ by $\ii \sqrt{|\det{\bf H}|}$ 
in Eq.\ (\ref{eq:mat_Xi-h-a_1d}),
such that $\Theta = \frac{\hbar \beta }{2} \sqrt{|\det{\bf H}|}$.
Differently from the elliptic case, note that 
${\rm Tr} {\bf \Xi}_{\text{H}} = 2\cos(\Theta) {\rm Tr}{\bf C} /(\hbar \beta)$ 
can take negative values,
which would imply ${\bf \Xi}_{\text{H}} \ngeq 0$.
Thus, to satisfy the criterion (\ref{eq:CondGKSL}), we should have
\begin{align*}
0 \le \Theta \le \tfrac{\pi}{2}
\,\,\, \text{or} \,\,\,
%\frac{4N+3}{2}\pi \le \Theta  \le \frac{4N +5}{2}\pi
\tfrac{\pi}{2} + (2N+1)\pi \le \Theta  \le \tfrac{\pi}{2} + 2(N+1)\pi  
\end{align*}
for $N=0,1,2...$, and
\begin{align*}
\left|\frac{ {\rm Tr}({\bf CH}) }{2\sqrt{\det({\bf CH})}} \tan\Theta \right| \le 1 \, .
\end{align*}
As an example,
scattering through a parabolic barrier is
described by the Hamiltonian
$\hat H = \tfrac{1}{2}H_{22} p^2 - \tfrac{1}{2}|H_{11}| q^2$
with $H_{11} > 0$, $H_{22} <0$, and $H_{12} =0$.
Although the \hy{CPTP} criterion is applicable, 
such a (quantum or classical) system will not reach thermal equilibrium 
due to the Hamiltonian not being bounded from below, 
as discussed in Sec.\ref{sec:CQoGBM}.

For the parabolic case, $\det{{\bf H}} =0$, we find that 
${\bf \Xi}_{\text{H}} = 2{\bf K}
-  \tfrac{\ii}{2} {\rm Tr}({\bf CH}) {\sf J}$, 
and the \hy{CPTP} condition becomes 
\[
\frac{1}{4} \frac{[\hbar \beta \, \tr{({\bf CH}})]^2}{\det{{\bf C}}} \le 1 \, ;
\]
the simplest example of such a system is the free particle Hamiltonian.

%%%%%%%%%%%%%%%%%%%%%%%%%%%%%%%%%%%%%%%%%%%%%%%%%%%%%%%%%%%%%%%%%%%%%%%%%%%%%%%%%%%%%%%%%
\subsection{Caldeira-Leggett Model}    
%%%%%%%%%%%%%%%%%%%%%%%%%%%%%%%%%%%%%%%%%%%%%%%%%%%%%%%%%%%%%%%%%%%%%%%%%%%%%%%%%%%%%%%%%
Our criterion is still applicable to a \hy{GTCL} equation 
which is not cast into the form of Eq.\ (\ref{eq:Q_Kramers}).
Let us consider the Caldeira-Leggett equation in a high-temperature environment \cite{caldeira1983}: 
\begin{align*}
\partial_t \hat \rho &=
\frac{i}{\hbar}  [ \hat \rho , \hat H_\text{o} +
\tfrac{1}{2} \gamma_\text{o} \{\hat q , \hat p\}]
- \frac{\zeta }{\hbar^2 \beta}[\hat q,[\hat q,\hat \rho]] \nonumber \\
&- \frac{i}{\hbar}\gamma_\text{o}[\hat q, \hat \rho \hat p] +
   \frac{i}{\hbar}\gamma_\text{o}[\hat p, \hat \rho \hat q] \, , %\label{eq:CLme}
\end{align*}
where $\hat H_\text{o}$ is a ``renormalized'' version of the original 
system Hamiltonian, 
$\gamma_\text{o}$ is a relaxation constant and $\zeta$ is a damping coefficient.

This equation is not expressed in the form of 
our quantum master equation (\ref{eq:Q_Kramers}), 
but can be mapped to the form of the \hy{GTCL} for 
the following identifications: 
\begin{align}
\begin{split}
{\bf \Xi}_\text{H} &= \frac{2\zeta}{\hbar \beta}
\left(
\begin{array}{cc}
1 & 0 \\
0 & 0
\end{array}
\right)
-\ii {\gamma_\text{o}}{\mathsf J}\, , 
%\end{aligned}
\\
{\bf \Xi}_\text{A} &= - \ii \gamma_\text{o}
                      \left( \begin{array}{cc}
                             0 & 1 \\
                             1 & 0
                             \end{array}    \right) \, ; 
\end{split} \nonumber
\end{align}
note that $\hat H_\text{eff} =
\hat H_\text{o} + \frac{\gamma_\text{o}}{2} \{\hat q , \hat p\}$. 
One can see that one of the eigenvalues of ${\bf \Xi}_\text{H}$ 
takes a negative value,
\begin{equation*}
\psi_{\pm} ({\bf \Xi}_\text{H}) 
= \frac{\zeta }{\hbar \beta} \pm
  \sqrt{\frac{\zeta^2}{\hbar^2 \beta^2} + \gamma_\text{o}^2 } \gtrless 0 \, ,
\end{equation*}
showing that the Caldeira-Leggett equation is not a \hy{GKSL} 
equation from Eq.\ (\ref{eq:CondGKSL}),
as expected \cite{BreuerPetr2002}.

Now we study the conditions for our master equation (\ref{eq:Q_Kramers}) 
to be reduced to this Caldeira-Leggett equation. 
To this end, we need to find matrices $\bf K$ and ${\mathsf S}_{\!\beta}$ 
satisfying Eq.\ (\ref{eq:mat_Xi}).
From Eq.\ (\ref{eq:mat_Xi-h-a}), one obtains the 
matrix $\bf \Xi$ of the Caldeira-Leggett model using the above matrices 
${\bf \Xi}_\text{H}$ and ${\bf \Xi}_\text{A}$.
%
%\[
%{\bf \Xi} = \tfrac{1}{2}({\bf \Xi}_\text{H} + {\bf \Xi}_\text{A}) =
%          \begin{pmatrix}
%            D_\text{o}/\hbar & -i \gamma_\text{o} \\
%             0& 0
%          \end{pmatrix}\, .
%\]
Thus the quantities satisfying the relation 
${\bf \Xi} = {\bf K}{\mathsf S}_{\!\beta}$ are defined 
if and only if  
\begin{align*}
\begin{split}%\label{eqn:para_CL}
\gamma_p &= \zeta \, {\rm sech}\Theta \, , \\
\gamma_\text{o} &= \zeta H_{22} \Theta^{-1} \tanh\Theta   \, ,
\end{split}
\end{align*}
together with $H_{12} = 0$ and $\gamma_q = 0$.
Since $\gamma_q = 0$, the Caldeira-Leggett equation is regarded 
as a special case of our quantum master equation (\ref{eq:Q_Kramers}), 
where the interactions with the heat bath do not modify the particle momenta
 --- this is the common perspective in standard Brownian motion \cite{risken}.
The nullity of $H_{12}$ means that the above identification works,  
for instance, for the harmonic oscillator and for the free-particle cases, 
where there are no contributions proportional to $\hat q \hat p$ or $\hat p \hat q$.
As discussed in Sec.\ref{sec:EC}, $\gamma_p$ is temperature-dependent. 

%%%%%%%%%%%%%%%%%%%%%%%%%%%%%%%%%%%%%%%%%%%%%%%%%%%%%%%%%%%%%%%%%%%%%%%%%%%%%%%%%%%%%%%%%
\subsection{Ordinary Kramers Equation}
%%%%%%%%%%%%%%%%%%%%%%%%%%%%%%%%%%%%%%%%%%%%%%%%%%%%%%%%%%%%%%%%%%%%%%%%%%%%%%%%%%%%%%%%%
The ordinary Kramers equation is obtained from the generalized Kramers 
equation in Eq.\ (\ref{eq:C_Kramers}) by setting $\gamma_{q_i} = 0, \forall i$.  
When Hamiltonians is bounded from below, 
its evolution by Eq.\ (\ref{eq:C_Kramers}) drives the system towards 
thermal equilibrium \cite{risken}. 
To show the quantization problem in the ordinary Kramers equation, 
it is sufficient to consider $n=1$.  

Setting $\gamma_q = 0$, we obtain
\[
\begin{aligned}
{\bf \Xi}_\text{H}& =
\frac{2\gamma_p}{\hbar \beta}\cosh\Theta
      \begin{pmatrix}
      1 & 0 \\ 0 & 0
      \end{pmatrix}
-  \ii \gamma_p H_{22}
\frac{\sinh\Theta}{2\Theta} {\sf J} \, , \\
{\bf \Xi}_\text{A} &= - \ii \gamma_p
\frac{\sinh\Theta}{ 2 \Theta}
                \begin{pmatrix}
                 2H_{12} & H_{22} \\
                  H_{22} & 0
                \end{pmatrix} ,
\end{aligned}
\]
from which we can calculate
\[
\det({\bf \Xi}_\text{H}) = 
- [2 H_{22} \gamma_p {\Theta}^{-1} \sinh(\Theta) ]^2\, .
\]
Consequently, the criterion (\ref{eq:CondGKSL}) is fulfilled only
for a Hamiltonian with $H_{22} = 0$.
This corresponds to the hyperbolic case where $\det {\bf H} = - H_{12}^2$  
and $\Theta = \tfrac{\hbar \beta}{2}\sqrt{|\det{\bf H}|}$.  
Since the hyperbolic Hamiltonian is unbounded, 
the system does not relax to equilibrium. 

%%%%%%%%%%%%%%%%%%%%%%%%%%%%%%%%%%%%%%%%%%%%%%%%%%%%%%%%%%%%%%%%%%%%%%%%%%%%%%%%%%%%%%%%%
%%%%%%%%%%%%%%%%%%%%%%%%%%%%%%%%%%%%%%%%%%%%%%%%%%%%%%%%%%%%%%%%%%%%%%%%%%%%%%%%%%%%%%%%%
\section{Application to network model for $n=2$  }  \label{sec:IO}   %%%%%%%%%%%%%%%%%%%%
%%%%%%%%%%%%%%%%%%%%%%%%%%%%%%%%%%%%%%%%%%%%%%%%%%%%%%%%%%%%%%%%%%%%%%%%%%%%%%%%%%%%%%%%%
%%%%%%%%%%%%%%%%%%%%%%%%%%%%%%%%%%%%%%%%%%%%%%%%%%%%%%%%%%%%%%%%%%%%%%%%%%%%%%%%%%%%%%%%%
Let us now consider two oscillators with equal masses and frequencies, 
which interact linearly with each other. Each oscillator, 
indexed by $i$, interacts with a heat bath at temperature $\beta_i$. 
This model, commonly referred to as the network model, has been extensively 
used to study heat conduction in stochastic and quantum thermodynamics
\cite{sekimoto,kosloff,volovich,cattaneo,basharov,nicacio7}. 
To simplify the notation, we redefine the position and momentum variables 
in such a way that both of them have the same dimension, $[q_i] = [p_j]$.
Then the Hamiltonian operator is defined by 
\begin{equation*} 
\hat{H} = \tfrac{\omega}{2}(\hat{p}_1^2 + \hat{p}_2^2 + \hat{q}_1^2 + \hat{q}_2^2)
  + \tfrac{\kappa}{2}(\hat{q}_1-\hat{q}_2)^2 \, ,
\end{equation*}
where $\kappa >0$ characterizes the interaction between the oscillators
and the dimensions of the parameters are such that 
$[\omega] = [\kappa] = [(\gamma_{p_i})^{-1}] = [(\gamma_{q_j})^{-1}]$.
%
% We apply this Hamiltonian operator for $n=2$ to the quantum master 
%equation (\ref{eq:Q_Kramers}).

The matrix ${\bf \Xi}$ in (\ref{eq:mat_Xi}) is calculated as 
\begin{align*}
{\bf \Xi} = \!
\begin{pmatrix}
\frac{\gamma_{p_1}}{\hbar\beta_1} ({\sf S}_{\beta_1})_{11} &
\frac{\gamma_{p_1}}{\hbar\beta_1} ({\sf S}_{\beta_1})_{12} &
\frac{\gamma_{p_1}}{\hbar\beta_1} ({\sf S}_{\beta_1})_{13} &
\frac{\gamma_{p_1}}{\hbar\beta_1} ({\sf S}_{\beta_1})_{14} \\
\frac{\gamma_{p_2}}{\hbar\beta_2} ({\sf S}_{\beta_2})_{21} &
\frac{\gamma_{p_2}}{\hbar\beta_2} ({\sf S}_{\beta_2})_{22} &
\frac{\gamma_{p_2}}{\hbar\beta_2} ({\sf S}_{\beta_2})_{23} &
\frac{\gamma_{p_2}}{\hbar\beta_2} ({\sf S}_{\beta_2})_{24} \\
\frac{\gamma_{q_1}}{\hbar\beta_1} ({\sf S}_{\beta_1})_{31} &
\frac{\gamma_{q_1}}{\hbar\beta_1} ({\sf S}_{\beta_1})_{32} &
\frac{\gamma_{q_1}}{\hbar\beta_1} ({\sf S}_{\beta_1})_{33} &
\frac{\gamma_{q_1}}{\hbar\beta_1} ({\sf S}_{\beta_1})_{34} \\
\frac{\gamma_{q_2}}{\hbar\beta_2} ({\sf S}_{\beta_2})_{41} &
\frac{\gamma_{q_2}}{\hbar\beta_2} ({\sf S}_{\beta_2})_{42} &
\frac{\gamma_{q_2}}{\hbar\beta_2} ({\sf S}_{\beta_2})_{43} &
\frac{\gamma_{q_2}}{\hbar\beta_2} ({\sf S}_{\beta_2})_{44}
\end{pmatrix} ,
\end{align*}
where 
\begin{align*}
({\sf S}_{\!\beta})_{ii} &=
\tfrac{1}{2} \cosh(\tfrac{\hbar \beta \omega}{2}) +
\tfrac{1}{2} \cosh(\tfrac{\hbar \beta \vartheta}{2})\nonumber \, ,\\
%\, , \,\,\, i = 1,2,3,4 \, ;\\
%
({\sf S}_{\!\beta})_{12} & =
({\sf S}_{\!\beta})_{21} =
\tfrac{1}{2} \cosh(\tfrac{\hbar \beta \omega}{2}) -
\tfrac{1}{2} \cosh(\tfrac{\hbar \beta \vartheta}{2})\nonumber \, ,\\
({\sf S}_{\!\beta})_{13} & = ({\sf S}_{\!\beta})_{24} =
-\tfrac{\ii}{2} \sinh(\tfrac{\hbar \beta \omega}{2})
- \tfrac{\ii\omega}{2\vartheta}\sinh(\tfrac{\hbar \beta \vartheta}{2})\nonumber \, ,\\
({\sf S}_{\!\beta})_{14} & = ({\sf S}_{\!\beta})_{23} =
-\tfrac{\ii}{2} \sinh(\tfrac{\hbar \beta \omega}{2})
+ \tfrac{\ii\omega}{2\vartheta}\sinh(\tfrac{\hbar \beta \vartheta}{2})\nonumber\, ,\\
({\sf S}_{\!\beta})_{34} &= ({\sf S}_{\!\beta})_{43} =
\tfrac{1}{2} \cosh(\tfrac{\hbar \beta \omega}{2}) -
\tfrac{1}{2} \cosh(\tfrac{\hbar \beta \vartheta}{2})\nonumber \, , \\
({\sf S}_{\!\beta})_{31} & =
 ({\sf S}_{\!\beta})_{42} =
\tfrac{\ii}{2} \sinh(\tfrac{\hbar \beta \omega}{2})
+ \tfrac{\ii\vartheta}{2\omega}\sinh(\tfrac{\hbar \beta \vartheta}{2})  \nonumber
\, ,\\
({\sf S}_{\!\beta})_{32} &= ({\sf S}_{\!\beta})_{41} =
\tfrac{\ii}{2} \sinh(\tfrac{\hbar \beta \omega}{2}) - 
\tfrac{\ii\vartheta}{2\omega}\sinh(\tfrac{\hbar \beta \vartheta}{2})  \, ,\nonumber
\end{align*}
and $\vartheta := \sqrt{\omega(\omega + 2 \kappa)}$.

\begin{figure}[ht!]
\includegraphics[width=\columnwidth]{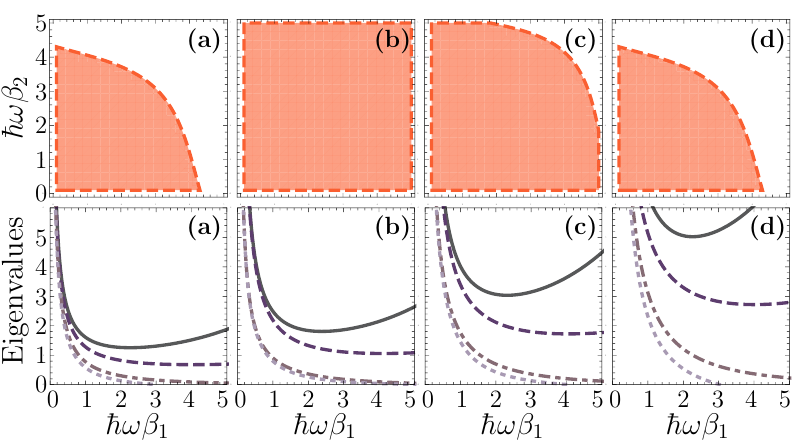}
\caption{CPTP Condition for the Network of Oscillators.
In the top panels, all eigenvalues in the shaded regions 
of the $\beta_1 \times \beta_2$ plane are non-negative. 
In the bottom panels, 
all four eigenvalues of ${\bf \Xi}_{\text{H}}$ are shown as 
functions of $\beta_1$ and $\beta_2=\beta_1$.
In all figures, 
$\gamma_{q_1} = \gamma_{q_2}$, 
$\gamma_{p_1} = \gamma_{p_2}$, 
and $\kappa/\omega = 1$.
For the remaining parameters, 
(a) $\gamma_{q_2} = \gamma_{p_2} = 1/(4\omega)$; 
(b) $\gamma_{q_2} = 1/(4\omega)$ and $\gamma_{p_2} = 1/(2\omega)$; 
(c) $\gamma_{q_2} = 1/(4\omega)$ and $\gamma_{p_2} = 3/(4\omega)$;  
(d) $\gamma_{q_2} = \gamma_{p_2} = 1/\omega$. }                  
\label{fig1}
\end{figure}

The eigenvalues of ${\bf \Xi}_{\text{H}}$ are numerically determined 
and their behaviors are shown in Fig.\ \ref{fig1}. 
In the top panels, 
all eigenvalues in the shaded regions of the $\beta_1 \times \beta_2$ plane 
are non-negative, showing that the time evolution of 
the quantum master equation (\ref{eq:Q_Kramers}) is a \hy{CPTP} map 
for the corresponding system parameters, 
according to the criterion (\ref{eq:CondGKSL}). 
In the bottom panels, all four eigenvalues of ${\bf \Xi}_{\text{H}}$ 
are shown as functions of $\beta_1$ for $\beta_2=\beta_1$.
These results suggest that it is possible to find appropriate parameters 
for which the quantum master equation (\ref{eq:Q_Kramers}) conforms to 
a \hy{CPTP} map for any temperature difference. 

Before concluding the discussion in this section, 
let us examine the classical limit of the matrix ${\bf \Xi}$,
\begin{align*}
{\bf \Xi} \approx
\begin{pmatrix}
\frac{\gamma_{p_1}}{\hbar\beta_1} & 0 &
-\tfrac{\ii}{2} \gamma_{p_1} \omega & 0 \\
0 & \frac{\gamma_{p_2}}{\hbar\beta_2} &
0 & -\tfrac{\ii}{2} \gamma_{p_2} \omega \\
\tfrac{\ii}{2} \gamma_{q_1} (\kappa+\omega) & -\tfrac{\ii}{2} \gamma_{q_1} \kappa &
\frac{\gamma_{q_1}}{\hbar\beta_1} & 0 \\
 -\tfrac{\ii}{2} \gamma_{q_2} \kappa & \tfrac{\ii}{2} \gamma_{q_2} (\kappa+\omega) &
0 & \frac{\gamma_{q_2}}{\hbar\beta_2}
\end{pmatrix} \, .
\end{align*}
It is then easy to see that the criterion is satisfied, 
${\bf \Xi}_{\text{H}} = {\bf K} >0$, 
which is the same behavior in the classical 
limit found in Sec.\ \ref{sec:LKE}.

%%%%%%%%%%%%%%%%%%%%%%%%%%%%%%%%%%%%%%%%%%%%%%%%%%%%%%%%%%%%%%%%%%%%%
%%%%%%%%%%%%%%%%%%%%%%%%%%%%%%%%%%%%%%%%%%%%%%%%%%%%%%%%%%%%%%%%%%%%

%%%%%%%%%%%%%%%%%%%%%%%%%%%%%%%%%%%%%%%%%%%%%%%%%%%%%%%%%%%%%%%%%%%%%%%%%%%%%%%%%%%%%%%%%
%%%%%%%%%%%%%%%%%%%%%%%%%%%%%%%%%%%%%%%%%%%%%%%%%%%%%%%%%%%%%%%%%%%%%%%%%%%%%%%%%%%%%%%%%
\section{Concluding Remarks} \label{sec:CRaP}                   %%%%%%%%%%%%%%%%%%%%%%%%%
%%%%%%%%%%%%%%%%%%%%%%%%%%%%%%%%%%%%%%%%%%%%%%%%%%%%%%%%%%%%%%%%%%%%%%%%%%%%%%%%%%%%%%%%%
%%%%%%%%%%%%%%%%%%%%%%%%%%%%%%%%%%%%%%%%%%%%%%%%%%%%%%%%%%%%%%%%%%%%%%%%%%%%%%%%%%%%%%%%%
In this paper, assuming that quantum master equations are derived through the canonical 
quantization of generalized Brownian motions \cite{KoiNic2023}, 
we established a criterion to ensure the complete positivity of the quantum master
equation (\ref{eq:Q_Kramers}). 
The scope of this paper is confined to general quadratic Hamiltonians, 
where the quantum master equation retains a quadratic form. 
We demonstrated that our quantum master equation can 
be transformed into the general form defined by Eq.\ (\ref{eq:QTCL}) by appropriately
selecting the matrix ${\bf \Xi}$. 
We then explored the criterion for this general form to be represented by 
a \hy{CPTP} map, as given by Eq.\ (\ref{eq:CondGKSL}). 

The criterion (\ref{eq:CondGKSL}) 
is developed from the \hy{GKSL} ``theorem'' \cite{breuer2016,rivas2012}, 
and hence it is state-independent, 
relying solely on the system Hamiltonian and interaction parameters.
When the criterion is applied to the entire set of 
one-particle quadratic Hamiltonians, 
we determine the parameters satisfying the requirements for a 
\hy{CPTP} evolution.
Alternatively, considering a different perspective, 
bosonic Gaussian channels represent \hy{CPTP} operations 
defined over the set of bosonic Gaussian states. 
A mapping between two Gaussian states will be a Gaussian channel 
if it satisfies the Heisenberg uncertainty principle \cite{serafini}, 
which corresponds to, in the context of the present paper, a \hy{CPTP} criterion.
In fact, the evolution governed by the \hy{QTCL} equation (\ref{eq:QTCL}), 
when constrained by the \hy{CPTP} criterion (\ref{eq:CondGKSL}), 
generates a one-parameter Gaussian channel \cite{Heinosaari}.
Consequently, the criterion (\ref{eq:CondGKSL}) 
is equivalent to the Heisenberg uncertainty principle as mentioned. 
However, in our case, it applies without requiring knowledge 
about the state of the environment.

While this outcome aligns with that derived in Ref.\ \cite{KoiNic2023} 
in the application to non-interacting harmonic oscillators,
the methodology presented here is systematic and 
significantly simpler than the approach employed in the referenced work.
Currently, our focus lies predominantly on the application to the network model, 
where the system is represented as an ensemble of harmonic oscillators 
interacting linearly with each other. 
The non-unitary part ${\cal D}^{(i)}_\text{Q}[\hat{\rho}]$ in Eq.\ (\ref{eq:Q_Kramers}) 
for such a system is influenced by the interaction terms in the Hamiltonian $\hat{H}$.
This influence results in our quantum master equation for 
the network being a global master equation. 
While the Lindblad operators in standard global equation of a network model 
are typically assumed to be proportional to 
the ladder operators that diagonalize $\hat{H}$,
the non-unitary part of Eq.\ (\ref{eq:Q_Kramers}) does not exhibit such 
a simple structure. 
Consequently, numerical calculations are required to 
determine the parameters that satisfy the CPTP criterion, 
differently from the standard global equation in the network model.
The results are summarized in Fig.\ \ref{fig1}, 
where the region of parameters that meet the CPTP map requirement is highlighted.        
Exactly speaking, however, the existence of a parameter set to ensure 
that our evolution forms a \hy{CPTP} map for any combination 
of two temperatures, $(\beta_1, \beta_2)$, has not yet been demonstrated. 
This is left as a future task.

% The results of this paper suggest that it is possible to formulate 
% a unified framework for stochastic thermodynamics and quantum thermodynamics 
% even for interacting models.
% For example, 
% given the system Hamiltonian defined by the network model, 
% our quantum master equation offers an alternative framework 
% for describing heat conduction --- a topic currently under intense debate.
% %
% Two distinct types of master equations, global and local, 
% have been used to study heat conduction. 
% The debate over which approach holds 
% more promise remains unresolved \cite{kosloff, volovich, cattaneo, basharov}. 
% Our equation belongs to the global type.
% %
% Differently from the standard global equation, 
% the classical limit of our quantum master equation is reduced to the 
% generalized Brownian motion of linearly coupled harmonic oscillators.
% Because of this, we can study how Fourier law of heat conduction, 
% observed in stochastic thermodynamics \cite{sekimoto}, 
% is affected by quantum effects.

\blue{Recently, the debates have focused on the desired thermodynamical attributes 
and properties of a general quantum master equation. 
Various criteria and approaches have been discussed in 
Refs.\cite{dann,soret,alicki2023}. 
In this context, new derivations and modifications 
have been proposed \cite{becker,abbruzzo,cattaneo,trushechkin}. 
The advantage of our theory lies in its ability to derive equations that satisfy the
CPTP condition while remaining consistent with the results of stochastic
thermodynamics.}

\blue{Our findings suggest that a unified framework for stochastic 
and quantum thermodynamics can be formulated, even for interacting models. 
For instance, given the system Hamiltonian defined by the network model, 
our quantum master equation offers an alternative framework for 
describing heat conduction. In studying this issue, the structure of 
local and global quantum master equations 
has been extensively discussed \cite{kosloff,volovich,cattaneo} 
to clarify their applicable domains, thermodynamic consistency, 
complete positivity, 
and so on \cite{becker,abbruzzo,cattaneo,trushechkin,basharov}). 
Our master equation is also global for a network of oscillators, 
but unlike the standard global master equation, it reduces to the 
corresponding generalized Brownian motion in the classical limit, 
enabling the study of Fourier's law of heat conduction with quantum effects.}

In contrast to the microscopic derivation of, for instance, 
a Bloch-Redfield equation, 
it should be noted that our approach is phenomenological, 
since our quantum master equation is derived by applying canonical quantization 
to a generalized model of Brownian motion, which is already a coarse-grained theory. 
Just as viscous coefficients in hydrodynamics are determined by external inputs, 
such as the Green-Kubo-Nakano formula, 
the coefficients $\gamma_{q_{i}}$ and $\gamma_{p_{i}}$ in Eq.\ (\ref{eq:Q_Kramers}),  
and also in Eq.\ (\ref{eq:C_Kramers}), 
should be determined from microscopic theories. 
It might be possible to apply the projection operator technique 
to derive the Green-Kubo-Nakano formula for these parameters, 
as demonstrated in Refs.\ \cite{zwanzig,koidekodama}, for instance.

Certain limitations on the applicability of our quantum master equation 
emerge when its derivation from a Brownian motion model becomes important.  
Specifically, Eq.\ (\ref{eq:C_Kramers}) is derived under the assumption of 
Markov processes, 
which presumes that the relaxation time scale 
of the bath degrees of freedom is significantly faster and 
well separated from the relaxation scale of the system. 
Out of these conditions, which could arise in extremely low temperature regions,
the validity of our quantum master equation may be compromised. 

Our theory is currently applied only to particle systems, 
but it should be applicable to quantum fields, as well. 
In quantum field theory, it is relatively easy to handle processes 
where the number of particles changes \cite{neidig}, 
making such an extension important for studying the effects of 
chemical potential in quantum thermodynamics. These studies are future challenges.

%%%%%%%%%%%%%%%%%%%%%%%%%%%%%%%%%%%%%%%%%%%%%%%%%%%%%%%%%%%%%%%%%%%%%%%%%%%%%%%%%%%%%%%%%
\begin{acknowledgments}%%%%%%%%%%%%%%%%%%%%%%%%%%%%%%%%%%%%%%%%%%%%%%%%%%%%%%%%%%%%%%%%%%
%%%%%%%%%%%%%%%%%%%%%%%%%%%%%%%%%%%%%%%%%%%%%%%%%%%%%%%%%%%%%%%%%%%%%%%%%%%%%%%%%%%%%%%%%

\noindent 
The authors are grateful for the helpful comments 
and suggestions provided by a referee. 
T.K. acknowledges the fruitful discussion with theory groups of the 
Institute for Theoretical Physics in the Johann Wolfgang Goethe University 
and the Frankfurt Institute for Advanced Studies (FIAS).
T.K. acknowledges the financial support by CNPq (No.\ 305654/2021-7) 
and the Fueck-Stiftung.
A part of this work has been done under the project INCT-Nuclear Physics
and Applications (No.\ 464898/2014-5);
F.N. is a member of the Brazilian National Institute of Science and Technology
for Quantum Information [CNPq INCT-IQ (465469/2014-0)].
\end{acknowledgments}

\appendix 

%%%%%%%%%%%%%%%%%%%%%%%%%%%%%%%%%%%%%%%%%%%%%%%%%%%%%%%%%%%%%%%%%%%%%%%%%%%%%%%%%%%%%%%%%
\section{Wigner-Wyel Calculus} \label{app:WWC}         %%%%%%%%%%%%%%%%%%%%%%%%%%%%%%%%%%
%%%%%%%%%%%%%%%%%%%%%%%%%%%%%%%%%%%%%%%%%%%%%%%%%%%%%%%%%%%%%%%%%%%%%%%%%%%%%%%%%%%%%%%%%

The action of the parity operator $\hat R_0 = \hat R_0^\dag$
on the column vector (\ref{eq:vec_op_x}),
$\hat R_0 \hat x \hat R_0= -\hat x$,
is a reflection over the origin \cite{sakuray}.
Using the Weyl operator in Eq.\ (\ref{eq:WyelMet_ops}), 
we define the displaced parity operator, 
or the reflection operator \cite{ozorio1998, grossmann-royer}, 
as $\hat R_x := \hat T_x \hat R_0 \hat T_x^\dag$. 
This operator is Hermitian ($\hat R_x = \hat R_x^\dag$) 
and unitary, making it an involutory operator ($\hat R_x^2 =\hat 1$), satisfying  
\[%begin{align}
\hat R_x \hat x \hat R_x^\dag = -\hat x + 2 x \, .
\]%end{align}

An arbitrary operator $\hat O$ acting on the Hilbert space of the
continuous-variable system of $n$ degrees of freedom
can be uniquely expanded as a linear combination of
the reflection operators \cite{ozorio1998}:
\[%\begin{equation} \label{eq:Wrep}
\begin{aligned}
\hat O &= \int \frac{d^{2n} x}{(\pi\hbar)^n}  \, O(x) \, {\hat R}_x.
\end{aligned}
\]%\end{equation}
Mathematically speaking,
the set $\{\hat R_x \,  | \, x \in {\mathbb R}^{2n}\}$
is a basis in the vector space of the operators.
The coefficient $O(x)$ of the expansion 
is the Wigner symbol of the operator $\hat O$ and given by
\begin{equation}\label{eq:Wsymb}
O(x)   = 2^n\tr\,( \hat O {\hat R}_x),
\end{equation}
by virtue of the following property \cite{ozorio1998}: 
\begin{equation*}
\tr\,({\hat R}_{x}{\hat R}_{x'}) = (\pi\hbar)^n\delta(x'-x) \, .
\end{equation*}
The trace in Eq.\ (\ref{eq:Wsymb}) on a coordinate basis
gives us the Wigner symbols \cite{grossmann-royer,ozorio1998}:
\[%\begin{equation} \label{eq:Wsymb2}
O(x) =
      \int_{\mathbb R^n} \!\!\! \!\!\! d^n \! q' \,
      \langle q + \textstyle \frac{1}{2} q'|
      \, \hat{O} \,
      | q - \textstyle \frac{1}{2} q' \rangle \,
      {\rm e}^{- \frac{\ii}{\hbar} p \cdot q'} \, .
\]%\end{equation}
The Wigner function $W(x)$ of a system is
(a normalized version of) the Wigner symbol of the
corresponding density operator $\hat \rho$ \cite{grossmann-royer,ozorio1998},
\begin{equation*} %\label{eq:WF}
W(x) := \frac{1}{(\pi\hbar)^{n}} \tr \left[ \hat \rho {\hat R}_x \right].
\end{equation*}

The Wigner representation is a suitable platform to perform
semiclassical approximations derived in Ref.\ \cite{ozorio1998} by
composing and expanding the symbols (\ref{eq:Wsymb}).
The Wigner symbol for the product of operators,
say $\hat O_{12} = \hat O_1 \Hat O_2$,
is obtained by the Groenewold rule \cite{ozorio1998}:
\begin{equation*}
O_{12}(x) = % \textrm{Tr}(\hat O_1 \hat O_2 \hat R_x) \\
{\rm e}^{\tfrac{\ii \hbar}{2}
           \partial_{x'} \cdot {\sf J} \partial_{x''} } \,
           O_1(x')|_{x'= x} O_2(x'')|_{x''= x}   \, .
\end{equation*}
Note that, for quadratic polynomials in $x$,
which is the case of a quadratic Hamiltonian in Eq.\ (\ref{eqn:quadra_hamil})
taken as the Wigner symbol of Eq.\ (\ref{eq:quad_Ham}),
the expansion of the above exponential ends in second order \cite{ozorio1998}.

%%%%%%%%%%%%%%%%%%%%%%%%%%%%%%%%%%%%%%%%%%%%%%%%%%%%%%%%%%%%%%%%%%%%%%%%%%%%%%%%%%%%%%%%%
%%%%%%%%%%%%%%%%%%%%%%%%%%%%%%%%%%%%%%%%%%%%%%%%%%%%%%%%%%%%%%%%%%%%%%%%%%%%%%%%%%%%%%%%%
\section{Quadratic Hamiltonians} \label{app:QH}        %%%%%%%%%%%%%%%%%%%%%%%%%%%%%%%%%%
%%%%%%%%%%%%%%%%%%%%%%%%%%%%%%%%%%%%%%%%%%%%%%%%%%%%%%%%%%%%%%%%%%%%%%%%%%%%%%%%%%%%%%%%%
%%%%%%%%%%%%%%%%%%%%%%%%%%%%%%%%%%%%%%%%%%%%%%%%%%%%%%%%%%%%%%%%%%%%%%%%%%%%%%%%%%%%%%%%%
Let us consider the evolution described by the unitary operator
\begin{equation} \label{eq:Unit_Quad_Ev}
\hat U_t =
            \hat T_{\xi} \, \hat M_{{\sf S}_t}\hat T_{\xi}^\dagger \,
            {\rm e}^{-\tfrac{\ii t}{\hbar} \varphi} \, ,
\end{equation}
where the Weyl operator,
a displacement operator in quantum optics \cite{ScullyZubairy},
and the metaplectic operator are, respectively,
defined by
\begin{align}
\begin{split}\label{eq:WyelMet_ops}
\hat T_\xi &:= {\exp}\!\left[\frac{\ii}{\hbar}  \hat x \cdot {\sf J} \xi \right]\, ,
\,\,\, \\
\hat M_{{\sf S}_t} &:= 
{\exp}\!\left[- \frac{\ii t}{2\hbar}  \hat x \cdot {\bf H} \hat x \right]\, .
\end{split}
\end{align}
Their actions in the operator $\hat x$ are
\begin{align}%\label{eq:CovRules}
\begin{split}
\hat T_\xi^\dag \hat x \hat T_\xi  &= \hat x + \xi \, , \\
\hat M_{{\sf S}_t}^\dag \hat x \hat M_{{\sf S}_t} &= {\sf S}_t \hat x\, ,
\end{split}\nonumber
%
%\hat M_{{\sf S}_t} \hat T_\xi \hat M_{{\sf S}_t}^\dag = \hat T_{ {\sf S}_t \xi}.
\end{align}
where $\xi \in \mathbb R^{2n}$ and the symplectic matrix ${\sf S}_t$ is
\begin{equation}\label{eq:Symp_S}
{\sf S}_t := \exp[{\mathsf J{\bf H}t}] \, .
\end{equation}
See \cite{littlejohn1986,gossonbook2006,nicacio16},
for instance.
From these covariance rules,
we obtain the Heisenberg evolution of the operator $\hat x$ as
\begin{equation} \label{eq:CovRules}
\hat U_t^\dag \hat x \hat U_t = {\sf S}_t (\hat x - \xi) + \xi \, .
\end{equation}
Note that for the real column vector 
$\xi := (\xi_1,...\,,\xi_{2n})^\top \in {\mathbb R}^{2n}$,  
the sum $\hat x'= \hat x + \xi$ 
should be interpreted as a column vector of operators 
with components $\hat x'_j = \hat x_j + \xi_j \hat 1_j$, 
where $\hat 1_j$ is the identity operator acting 
on the Hilbert space associated to the 
$(j \, {\rm mod} \, n)^\text{th}$ degree of freedom.

The unitary operator in Eq.\ (\ref{eq:Unit_Quad_Ev}) is a member of
the inhomogeneous metaplectic group \cite{littlejohn1986,gossonbook2006} ,
a unitary subgroup generated by quantum quadratic Hamiltonians:
\begin{equation*}
\hat H := \ii\hbar \hat U_t^{-1}\frac{d\hat U_t}{dt}  =
\frac{1}{2} (\hat x-\xi) \cdot {\bf H}  (\hat x-\xi) + \varphi\hat 1\, ,
\end{equation*}
which is the symmetric quantization of the Hamiltonian
in Eq.\ (\ref{eqn:quadra_hamil}) and is the quantum operator in Eq.\ (\ref{eq:quad_Ham}).

Performing a Wick rotation $t \longmapsto -{\ii} \hbar\beta_k/2$ with
$\beta_k \in {\mathbb R}$ in the unitary evolution (\ref{eq:Unit_Quad_Ev}),
one obtains $\hat U_t \,\, \longmapsto \,\, \exp[ \tfrac{1}{2} \beta_k \hat{H}]$.
Consequently, from (\ref{eq:CovRules}),
\begin{equation}\label{eq:wick}
\hat U_t^\dag \hat x \hat U_t \longmapsto
\hat y := {\sf S}_{\! \beta_k} (\hat x - \xi ) + \xi,
\end{equation}
where, from Eq.\ (\ref{eq:Symp_S}),
\begin{equation}\label{eq:WRSbeta}
{\sf S}_{t} \longmapsto {\sf S}_{\!\beta_k}
:= {\exp}[- \ii \hbar \beta_k {{\sf J}{\bf H} }/2] \, ,
\end{equation}
which is a complex symplectic matrix belonging to the
Wick rotated symplectic group \cite{nicacio16}, {\it i.e.},
\begin{equation} \label{eq:symSb}
\begin{aligned}
&{\sf S}_{\! \beta_k}^\top {\sf J} {\sf S}_{\! \beta_k} = {\sf J} \, ,\\
&{\sf S}_{\! \beta_k}^{-1} = {\sf S}_{-\beta_k} = {\sf S}_{\! \beta_k}^\ast \, .
\end{aligned}
\end{equation}
Note that $\hat{y}^\dag = {\sf S}_{\! \beta_k}^\ast (\hat x - \xi ) + \xi$,
for $\hat y$ in Eq.\ (\ref{eq:wick}).

%%%%%%%%%%%%%%%%%%%%%%%%%%%%%%%%%%%%%%%%%%%%%%%%%%%%%%%%%%%%%%%%%%%%%%%%%%%%%%%%%%%%%%%%%
\section{Derivation of the QTCL Equation} \label{app:QuadGen}            %%%%%%%%%%%%%%%%
%%%%%%%%%%%%%%%%%%%%%%%%%%%%%%%%%%%%%%%%%%%%%%%%%%%%%%%%%%%%%%%%%%%%%%%%%%%%%%%%%%%%%%%%%
The most general linear form of ${\cal D}[\hat{\rho}]$ 
for a quadratic dynamics is given by 
\begin{eqnarray*} %\label{eq:TCQD}
{\mathcal D}[\hat \rho] = & - & \frac{1}{\hbar}\sum_{i,j=1}^{2n} 
({\bf L}_{ij} \hat x_i \hat x_j \hat \rho + 
{\bf M}_{ij} \hat x_i \hat \rho \hat x_j +
{\bf N}_{ij} \hat x_i \hat x_j \hat \rho ) \nonumber \\
& - & \frac{1}{\hbar} \sum_{i=1}^{2n} 
(\alpha_i \hat x_i \hat \rho + \beta_i \hat \rho \hat x_i) + c \hat \rho  \, ,
\end{eqnarray*}
where ${\bf L}$, ${\bf M}$, and ${\bf N}$ 
are arbitrary $2n \times 2n$ complex matrices,  
$\alpha, \beta \in {\mathbb C}^{2n}$ are arbitrary complex vectors, 
and $c \in {\mathbb C}$ is an arbitrary complex number. 

To ensure the trace preservation, 
$\tr {\mathcal D}[\hat \rho] = 0$, one finds 
\[
{\bf L} + {\bf M}^\top + {\bf N} = 0 \,,  \,\,\, 
\alpha + \beta = 0 \,,  \,\,\,
c = 0 \, ,
\]
while Hermiticity, 
${\mathcal D}^\dagger = {\mathcal D}$, 
requires 
\[
{\bf L} = {\bf N}^\dagger  \,,  \,\,\, 
{\bf M} = {\bf M}^\dagger  \,,  \,\,\,
\alpha = \beta^\ast \,.
\]
These constrains are solved by a complex 
$2n \times 2n$ matrix $\bf \Xi$ and a {\it real} vector $\eta \in \mathbb R^{2n}$. 
Indeed, one can write
\begin{align}
\begin{split}
{\bf L} &= {\bf \Xi}^\top \, , \\
{\bf M} &= - ({\bf \Xi} +{\bf \Xi}^\dagger) \, , \\
{\bf N} &= {\bf \Xi}^\ast \, ,\\
\alpha &= \beta^\ast = 2 \ii {\rm Im}({\bf \Xi}) \eta \, .  
\end{split} \nonumber 
\end{align}
After expanding all the parenthesis in Eq.\ (\ref{eq:QTCL}), 
it is easy to confirm that the latter is equivalent to Eq.\ (\ref{eq:QTCL}). 
The above discussion is still applicable when matrices and vectors are time-dependent.

We further discuss the reexpression of 
our quantum master equation (\ref{eq:Q_Kramers}) 
in the form of the \hy{QTCL} equation.
Using the notation in Eq.\ (\ref{eq:vec_op_x}),
the dissipative part of Eq.\ (\ref{eq:Q_Kramers}) may be reorganized as
\begin{eqnarray}\label{eq:Diss_Q_Kram}
\sum_{k=1}^n \mathcal{D}^{(k)}_\text{Q}[\hat \rho] =
&-&\frac{1}{{\hbar}}\sum_{i,j=1}^{2n} {\bf K}_{ij}
(  \hat \rho \, \hat{y}_i \hat x_j
 + \hat x_i \hat{y}_j^\dag \hat \rho)  \nonumber \\
&+&\frac{1}{{\hbar}}\sum_{i,j=1}^{2n} {\bf K}_{ij}
(  \hat x_i \hat \rho \hat{y}_j +
   \hat{y}_i^\dag \hat \rho \hat x_j ) \, .
\end{eqnarray}
for a generic Hamiltonian $\hat H$. 
In the above equation, ${\bf K}$ is defined in Eq.\ (\ref{eq:mat_K}) and
\begin{equation}\label{eq:vec_y}
\hat{y}_i := {\rm e}^{ \tfrac{1}{2} \beta_i \hat{H}} \hat x_i \,
                       {\rm e}^{-\tfrac{1}{2} \beta_i \hat{H}}  
\end{equation}
for $i =1,...,2n$ and with $\beta_{j+n} = \beta_j$ for $j = 1,...,n$. 
%
%Note that  
%{${\bf K} = {\hbar}^{-1}{\sf J}{\bf D}{\sf J}^\top$ 
%for ${\bf D}$ in Eq.\ (\ref{eq:matrices_CD}) and $\sf J$ in Eq.\ (\ref{eq:mat_J}).}
%

For the quadratic Hamiltonian (\ref{eq:quad_Ham}), 
the operator vector in Eq.\ (\ref{eq:vec_y}) is given in Eq.\  (\ref{eq:wick}) 
and 
\[
\hat{y}_i^\dag  = {\rm e}^{ - \tfrac{1}{2}\beta_i \hat{H}} \hat x_i \,
{\rm e}^{ \tfrac{1}{2}\beta_i \hat{H}} = {\sf S}_{\! \beta_i}^\ast (\hat x - \xi ) + \xi \, .
\]
Inserting these in Eq.\ (\ref{eq:Diss_Q_Kram}),
we found a \hy{QTCL} master equation (\ref{eq:QTCL}), 
where the matrix $\bf \Xi$ is given by Eq.\ (\ref{eq:mat_Xi}) and $\eta = \xi$.

% we can show that ${\sf S}_{\! \beta_i}$ in Eq.\ (\ref{eq:wick}) 
% is given by Eq.\ (\ref{eq:mat_KS}) with $\beta_{i+n} = \beta_i$ for $i = 1,...,n$.
%\[
%\hat y_i = \sum_{j=1}^{2n}({\sf S}_{\beta_i})_{ij} (\hat x_j - \xi_j) + \xi_i \, ,
%\]
%where ${\sf S}_{\beta_i}$ is given in Eq.\ (\ref{eq:mat_KS})
%and $\beta_{i+n} = \beta_i$ for $i = 1,...,n$.
%

%%%%%%%%%%%%%%%%%%%%%%%%%%%%%%%%%%%%%%%%%%%%%%%%%%%%%%%%%%%%%%%%%%%%%%%%%%%%%%%%%%%%%%%%%
\section{Uniqueness of the Matrix ${\bf \Xi}_\text{H}$} \label{app:non-unique} %%%
%%%%%%%%%%%%%%%%%%%%%%%%%%%%%%%%%%%%%%%%%%%%%%%%%%%%%%%%%%%%%%%%%%%%%%%%%%%%%%%%%%%%%%%%%
Normally, a quantum master equation is expressed as 
the sum of unitary and non-unitary generators of the evolution, 
this decomposition is however not unique.
As discussed in Refs.\ \cite{colla,NicMaia}, 
the \hy{GTCL} equation (\ref{eq:GTCL}) for general Lindblad operators
is invariant under the following transformation:  
\begin{equation}\label{eq:G-gauge}
\begin{aligned}
\hat L_\mu' &:=
\sum_{\nu =1}^{2n} 
{U}_{\mu\nu} \hat L_\nu + \varsigma_\mu \qquad (\mu = 1,...,2n), \\
\hat H'_{\rm eff} &:= \hat H_{\rm eff} + 
          \sum_{\mu,\nu =1}^{2n}
          \frac{ {g}_{\mu\nu} }{2\ii}
          (\varsigma_\mu^\ast {U}_{\nu\kappa}\hat L_\kappa -
           \varsigma_\mu      {U}_{\nu\kappa}^\ast\hat L_\kappa^\dagger),
\end{aligned}
\end{equation}
where $U$ is a complex $2n \times 2n$ unitary matrix satisfying $U g U^\dagger = g$, 
and $\varsigma_\mu$ are complex numbers. 
Reexpressing Eq.\ (\ref{eq:GTCL}) using the transformed operators, 
$\hat L_\mu' $ and $\hat H'_{\rm eff}$, 
changes the unitary and non-unitary parts of the \hy{GTCL} equation. 
Choosing $g_{\mu\nu} = \delta_{\mu\nu}$, 
the \hy{GTCL} equation is reduced to the \hy{GKSL} equation. 
This nonuniqueness for the \hy{GKSL} equation is already pointed out 
in the seminal \hy{GKSL} works \cite{GKSL} and similar transformations 
are reported in other descriptions of open systems.  
For instance, although the Bloch-Redfield equations (and generalizations) 
are considered in Ref.\ \cite{davidovic},   
the prescription follows the same structure of above transformation: 
a change in the Lindblad operators is compensated 
by a modification of the Hamiltonian, or {\it vice-versa}.

As shown in Sec.\ \ref{sec:GKSLME}, the \hy{QTCL} equation (\ref{eq:QTCL}) 
is reduced to the \hy{GTCL} equation (\ref{eq:GTCL}) 
with $\hat L_\mu = \lambda_\mu \cdot (\hat x - \eta )$.
Applying Eq.\ (\ref{eq:G-gauge}), these Lindblad operators become 
\begin{equation*}
\hat L'_\mu = {U}_{\mu\nu} \lambda_\nu \cdot (\hat x - \eta ) + \varsigma_\mu \, , 
\end{equation*} 
but the matrix ${\bf \Xi}_\text{H}$ defined by Eq.\ (\ref{eq:mat_Xih}) is invariant:
\begin{equation*}%\label{eq:mat_Xih_trans}
{\bf \Xi}'_\text{H} = {U}{\bf \Xi}_\text{H}{U}^\dagger = {\bf \Xi}_\text{H} \, .
\end{equation*}
%
%since $U g U^\dagger = g$.  
%
Noteworthy, 
neither the Hamiltonian $\hat H_{\rm eff}$
nor the matrices ${\bf \Xi}$ and ${\bf \Xi}_\text{A}$ in Eq.\ (\ref{eq:mat_Xi-h-a})
are invariant for the same transformation (\ref{eq:G-gauge}). 

%%%%%%%%%%%%%%%%%%%%%%%%%%%%%%%%%%%%%%%%%%%%%%%%%%%%%%%%%%%%%%%%%%%%%%%%%%%%%%%%%%%%%%%%%
\section{Quantum Detailed Balance Conditions}  \label{app:QDBC} %%%%%%%%%%%%%%%%%%%
%%%%%%%%%%%%%%%%%%%%%%%%%%%%%%%%%%%%%%%%%%%%%%%%%%%%%%%%%%%%%%%%%%%%%%%%%%%%%%%%%%%%%%%%%
For a \hy{GKSL} equation written as Eq.\ (\ref{eq:GTCL}), 
the quantum detailed balance conditions \cite{QDBCfin} 
are expressed, for $\mu = 1, ..., n$, as   
%\label{Eq:DetBal}
\begin{align}
\begin{split}\nonumber
&{\text{\it (i)}} \,\,\,  [ \hat H , \hat H_\text{eff} ] = 0; \\    
&{\text{\it (ii)}} \,\,\, \hat U^\dagger \hat L_{\mu} \hat U  = 
              {\rm e}^{-i \omega_\mu t } \hat L_{\mu}  \, , \,\,\, 
              \hat U :=  {\rm e}^{-\frac{i}{\hbar} \hat H t} \, , \quad \\  
&{\text{\it (iii)}} \,\,\, \hat L_{\mu + n} =
{\rm e}^{ -\tfrac{1}{2}\beta \hbar \omega_\mu} \hat L_{\mu}^\dag \, . 
\end{split}
\end{align}
These necessary and sufficient conditions relate the Bohr 
frequencies $\omega_\mu$ of the system Hamiltonian $\hat H$ 
with the Lindblad operators in order to asymptotically drive 
the system to the equilibrium state. 
The \hy{GKSL} equation in Eq.\ (\ref{eq:Optics_ME}) is one example: 
the operators in Eq.\ (\ref{eq:OHLind}) and 
a harmonic oscillator Hamiltonian are detailed balanced.
In this particular case $\hat H = \hat H_\text{eff}$, 
and the asymptotic state of Eq.\ (\ref{eq:Optics_ME}) is 
a thermal Gibbs state $\sim\exp[-\beta \hat H]$ \cite{BreuerPetr2002}. 
These conditions for \hy{GKSL} equation, 
written as the \hy{GTCL} equation (\ref{eq:GTCL}), 
for a quadratic Hamiltonian and linear Lindblad operators are thoroughly 
studied in Ref.\ \cite{toscano2022}.  

For the quadratic Hamiltonian in Eq.\ (\ref{eq:Heff}), 
using $\hat H$ in Eq.\ (\ref{eq:quad_Ham}) and the canonical commutation relation 
in Eq.\ (\ref{eq:ComRel}), we determine \cite{littlejohn1986,gossonbook2006,nicacio16}
\[
\begin{aligned}
[ \hat H , \hat H_\text{eff} ] & =  
\frac{\ii}{4} [ (\hat x-\xi) \cdot {\bf H}  (\hat x-\xi) ,  
              (\hat x-\xi) \cdot {\bf \Xi}^\ast_\text{A} (\hat x-\xi) ] \\
& = - \frac{\hbar}{2} (\hat x-\xi) \cdot 
( {\bf H}{\sf J}{\bf \Xi}^\ast_\text{A} - {\bf \Xi}^\ast_\text{A}{\sf J}{\bf H}) 
(\hat x-\xi)  \, , 
\end{aligned}
\]
where $\xi = \eta$. 
Consequently, the condition ({\it i}) can be written as 
$ (-{\sf J}{\bf H})^\top {\bf \Xi}_\text{A} = {\bf \Xi}_\text{A}{\sf J}{\bf H}$, 
which is equivalent to 
\begin{equation*}
%\label{eq:Quad(i)}
{\sf S}_{t}^\top  {\bf \Xi}_\text{A} {\sf S}_{t} = {\bf \Xi}_\text{A} \, , 
\,\,\, 
{\sf S}_{t} = {\rm e}^{{\sf J} {\bf H} t } \, .
\end{equation*}

For a linear Lindblad operator $\hat L_\mu = \lambda_\mu \cdot (\hat x - \xi)$, 
using Eqs.\ (\ref{eq:Symp_S}) and (\ref{eq:CovRules}), 
the condition {({\it ii})} is thus expressed as an eigenvalue equation: 
\begin{equation}\label{eq:Quad(ii)}
{\sf S}_{t}^\top \lambda_\mu = {\rm e}^{-i \omega_\mu t } \lambda_\mu   \, ,  
\end{equation}
because 
\[
\hat U^\dag \hat L_{\mu} \hat U = \lambda_\mu \cdot ( \hat U^\dag \hat x \hat U - \xi) 
= {\sf S}_t^\top \lambda_\mu \cdot ( \hat x - \xi) \, .
\]
The condition {({\it iii})} immediately reads
\begin{equation}\label{eq:Quad(iii)}
\lambda_{\mu + n} = 
{\rm e}^{ -\tfrac{1}{2}\beta \hbar \omega_\mu}\lambda_{\mu}^\ast   \, . 
\end{equation}
From Eq.\ (\ref{eq:Quad(iii)}), the matrix in Eq.\ (\ref{eq:mat_Xih}) for 
$g_{\mu\nu} = \delta_{\mu\nu}$ becomes 
\begin{equation*}
%\label{eq:mat_XihDBC}
{\bf \Xi}_\text{H} = 
\sum_{\mu = 1}^{2n} \lambda_\mu \lambda_\mu^\dag = 
\sum_{\mu = 1}^{n} \left( \lambda_\mu \lambda_\mu^\dag + 
                          {\rm e}^{ -\beta \hbar \omega_\mu} 
                          \lambda_\mu^\ast \lambda_\mu^\top \right) \, . 
\end{equation*}
Using Eq.\ (\ref{eq:Quad(ii)}), it is then straightforward to show that
${\sf S}_{t}^\top  {\bf \Xi}_\text{H} {\sf S}_{t} = {\bf \Xi}_\text{H}$.  
Finally, from the definition in Eq.\ (\ref{eq:mat_Xi-h-a}), 
we find a necessary condition for quantum detailed balance of 
a quadratic \hy{GKSL} equation (\ref{eq:GTCL})
\[
{\sf S}_{t}^\top  {\bf \Xi} {\sf S}_{t} = {\bf \Xi} \, .
\] 
However, it will be more useful for our purposes 
to perform the Wick rotation (\ref{eq:WRSbeta}) in above relations to find
\begin{equation}\label{eq:invXi}
{\sf S}_{\beta}^\top  {\bf \Xi}_\text{H} {\sf S}_{\beta} = {\bf \Xi}_\text{H} \, , 
\,\,\,
{\sf S}_{\beta}^\top  {\bf \Xi}_\text{A} {\sf S}_{\beta} = {\bf \Xi}_\text{A} \, , 
\,\,\, 
{\sf S}_{\beta}^\top  {\bf \Xi} {\sf S}_{\beta} = {\bf \Xi} \, . 
\end{equation}

To require our master equation to describe thermal relaxation, 
the matrix in Eq.\ (\ref{eq:mat_Xi}) becomes 
${\bf \Xi} = {\bf K} {\sf S}_{\beta}$, since all the temperatures are equal. 
Applying relations in Eq.\ (\ref{eq:invXi}) for this matrix ${\bf \Xi}$ 
and using the properties of ${\sf S}_{\beta}$ in Eq.\ (\ref{eq:symSb}), 
we find 
\[
{\bf \Xi}_\text{A} = 0 \,, \,\,\,  
{\bf \Xi}_\text{H} = 2 {\bf \Xi} = 2 {\bf K} {\sf S}_{\beta}\, ,  
\]
which are necessary conditions for our master equation describe 
a thermal relaxation process.
From our condition (\ref{eq:CondGKSL}), 
the matrix ${\sf S}_{\beta} = \exp[-i \hbar \beta {\sf J}{\bf H}/2]$ 
must be positive semidefinite and, then, 
the Hamiltonian matrix ${\sf J}{\bf H}$ must have purely imaginary eigenvalues. 
Such Hamiltonian matrices are called elliptic \cite{nicacio16} 
and include harmonic oscillators and linearly coupled oscillators, 
both with ${\bf H} > 0$. 

%%%%%%%%%%%%%%%%%%%%%%%%%%%%%%%%%%%%%%%%%%%%%%%%%%%%%%%%%%%%%%%%%%%%%%%%%%%%%%%%%%%%%%%%%
%%%%%%%%%%%%%%%%%%%%%%%%%%%%%%%%%%%%%%%%%%%%%%%%%%%%%%%%%%%%%%%%%%%%%%%%%%%%%%%%%%%%%%%%%

%%%%%%%%%%%%%%%%%%%%%%%%%%%%%%%%%%%%%%%%%%%%%%%%%%%%%%%%%%%%%%%%%%%%%%%%%%%%%%%%%%%%%%%%%
%%%%%%%%%%%%%%%%%%%%%%%%%%%%%%%%%%%%%%%%%%%%%%%%%%%%%%%%%%%%%%%%%%%%%%%%%%%%%%%%%%%%%%%%%
%%%%%%%%%%%%%%%%%%%%%%%%%%%%%%%%%%%%%%%%%%%%%%%%%%%%%%%%%%%%%%%%%%%%%%%%%%%%%%%%%%%%%%%%%
\end{document}